\documentclass[preprint,12pt,numbers,nopreprintline]{elsarticle}

\usepackage{amsmath,amssymb}
\usepackage{graphicx}
\usepackage{booktabs}
\usepackage{multirow}
\usepackage{array}
\usepackage{tabularx}
\usepackage{threeparttable}
\usepackage{xcolor}
\usepackage{lineno}
\usepackage[hidelinks,pdfusetitle=false]{hyperref}
\usepackage{enumitem}
\usepackage{caption}
\usepackage{subcaption}
\usepackage{float}
\usepackage{pdflscape}
\usepackage[pass]{geometry}
\usepackage[section]{placeins}
\usepackage{etoolbox}

\patchcmd{\pprintMaketitle}{\vskip18pt}{\vskip12pt}{}{}
\patchcmd{\pprintMaketitle}{\vskip36pt}{\vskip12pt}{}{}
\patchcmd{\pprintMaketitle}{\clearpage}
  {\clearpage\enlargethispage{2\baselineskip}}{}{}
\graphicspath{{./}}
\biboptions{numbers,square,sort&compress}

\newcommand{\ModPresent}{\begingroup\setlength{\unitlength}{1pt}\begin{picture}(3.2,3.2)\put(1.6,1.6){\circle*{2.6}}\end{picture}\endgroup}
\newcommand{\ModAbsent}{\begingroup\setlength{\unitlength}{1pt}\begin{picture}(3.2,3.2)\linethickness{0.3pt}\put(1.6,1.6){\circle{2.6}}\end{picture}\endgroup}

\newcommand{\StateOneDice}{10.25}

\newcommand{\StateFourDice}{58.75}
\newcommand{\StateOneEntropy}{0.974}

\newcommand{\StateFourEntropy}{0.729}
\newcommand{\StateOneMaxP}{0.334}

\newcommand{\StateFourMaxP}{0.648}

\newcommand{\StateTwoMargin}{1.163}
\newcommand{\StateThreeMargin}{1.074}
\newcommand{\StateFourMargin}{0.938}
\newcommand{\StateOneTokenCos}{-0.286}

\newcommand{\StateFourTokenCos}{0.189}

\hypersetup{pdfauthor={},pdftitle={Progressive multi-scale tumor-state memory for incomplete multimodal brain tumor segmentation}}
\begin{document}
\begin{frontmatter}

\title{Progressive multi-scale tumor-state memory for incomplete multimodal brain tumor segmentation}

\author[inst1]{Milad Rostamian}
\ead{milad\_rostamian@comp.iust.ac.ir}
\author[inst1]{Mohammadreza Mohammadi}
\ead{mrmohammadi@iust.ac.ir}
\author[inst1]{Mohsen Soryani\corref{cor1}}
\ead{soryani@iust.ac.ir}
\cortext[cor1]{Corresponding author: Mohsen Soryani.}
\affiliation[inst1]{organization={School of Computer Engineering, Iran University of Science and Technology (IUST)}, addressline={Hengam Street, Resalat Square}, postcode={13114-16846}, city={Tehran}, country={Iran}}

\begin{abstract}
Brain tumor segmentation uses complementary magnetic resonance imaging
(MRI) sequences, but some sequences may be unavailable in practice.
Their absence changes the information available for segmenting
different tumor regions. Fusion methods can support different input
subsets. However, fusion at several scales does not by itself specify
how tumor evidence is retained and updated to guide fusion at later
stages. We propose a progressive multi-scale tumor-state memory
network (PMTM) for segmentation from any non-empty subset of T1ce,
T1, FLAIR, and T2. PMTM maintains a compact state for the current
subject across the encoder stages. At each stage, we use assignment
maps to pool spatial features into ordered evidence vectors for
each available modality. We average these vectors across the
available modalities, and a learned gate combines this evidence
with the previous state. The updated state guides residual fusion
at that stage and is used in the state update at the next stage.
A modality-token mixer also models interactions among the available
modalities at each stage. We apply semantic supervision to the
assignment maps at all four stages.
We evaluate PMTM on BraTS2018, BraTS2020, and BraTS2023 GLI across
all 15 non-empty modality subsets. Averaged over the three tumor
regions and the 15 subsets, PMTM achieves Dice scores of 75.31\%,
77.54\%, and 80.47\% on BraTS2018, BraTS2020, and BraTS2023 GLI,
respectively. These scores are higher than those of RobustSeg,
RFNet, and mmFormer in our experiments on each dataset. Component ablations and
internal state analysis are performed on BraTS2020.
\end{abstract}


\end{frontmatter}


\section{Introduction}
\label{sec:introduction}

Glioma segmentation on MRI supports diagnosis and treatment
planning \citep{jyothi2023review}.
The tumor regions considered here are the whole tumor (WT),
tumor core (TC), and enhancing tumor (ET)
\citep{menze2015brats,bakas2017advancing}.
To identify these regions, multimodal brain MRI commonly combines
contrast-enhanced T1-weighted (T1ce), T1-weighted (T1),
fluid-attenuated inversion recovery (FLAIR), and T2-weighted (T2)
images \citep{menze2015brats,bakas2017advancing}.
These sequences provide complementary information about the
tumor regions. T1ce is especially informative for TC and ET,
while FLAIR and T2 provide information about edema and WT
\citep{ding2021rfnet,xiao2025mpdfunet}.

Many multimodal segmentation models are designed to receive a complete
set of MRI sequences \citep{zhou2023missingreview}.
However, differences in acquisition protocols, image corruption,
resource constraints, and restrictions on contrast administration
can leave one or more sequences unavailable
\citep{zhou2023missingreview}.
The resulting loss of information affects tumor regions differently
\citep{ding2021rfnet,xiao2025mpdfunet}.
Segmentation from incomplete inputs therefore requires a fusion
mechanism that works with the available subset and accounts for
the information supplied by each remaining sequence.

Existing approaches address this problem in different ways.
Latent aggregation constructs a common representation from different
modality subsets \citep{havaei2016hemis,dorent2019uhved}.
Region-aware fusion accounts for how each sequence contributes
to individual tumor regions \citep{ding2021rfnet}.
Transformer modules learn interactions between modality
representations \citep{zhang2022mmformer,shi2024m2ftrans}.
Our earlier method, TG-BWSA, exchanges modality features between
encoder stages through a shared adapter
\citep{rostamian2026tgbwsa}.
These mechanisms determine how the available features are combined.
A related question is how evidence about tumor regions can be
accumulated across encoder stages and used to guide fusion
at later stages.

Memory mechanisms reuse information during feature extraction
and decoding \citep{wang2023memorynet,zhang2026hmdm}.
In multimodal brain tumor segmentation, MSM-Seg uses memory
to model relationships between modalities and across slices
\citep{luo2025msmseg}.
These studies motivate the use of retained context.
The role of memory depends on what is stored and where
the network uses it.
Here, we investigate a tumor state built from the available MRI
sequences that accumulates tumor evidence across encoder stages
and directly guides modality fusion.

We propose PMTM, a progressive multi-scale tumor-state memory network.
PMTM maintains a tumor state, an ordered set of feature vectors
called state slots.
At each encoder stage, assignment maps pool features from each
available modality into evidence vectors.
We average corresponding evidence vectors across the available
modalities and use them to update the state.
The updated state guides fusion at the current stage and is used
in the state update at the next stage.
Fusion uses the current modality features together
with the state accumulated from preceding stages.
At the beginning of each forward pass, the state is initialized
from a learned tensor. State updates use evidence from the
available sequences of the current subject.

The contributions of this work are summarized below.

\begin{enumerate}[leftmargin=*]

\item We introduce a progressive tumor-state path for segmentation
from incomplete multimodal MRI.
Ordered state slots accumulate evidence from the available
MRI sequences across all four encoder stages.

\item We apply semantic supervision to the assignment maps
at all four encoder stages.

\item We introduce residual fusion guided by the updated tumor state.
A modality-token mixer also models interactions among the available
modalities at each encoder stage.

\item We evaluate PMTM on BraTS2018, BraTS2020, and BraTS2023 GLI
across all 15 non-empty modality subsets.
Component ablations on BraTS2020 assess the tumor-state path,
cross-stage state propagation, semantic supervision, and
modality-token interaction.
Internal analyses examine the learned assignment maps and state slots.

\end{enumerate}

\section{Related work}
\label{sec:related_work}

\subsection{Fusion in complete multimodal brain tumor segmentation}

Methods for complete inputs differ in when MRI sequences are combined.
\citet{pereira2016brain} processed the four sequences jointly with
a patch-based CNN.
Separate encoders allow fusion to adjust modality contributions
after feature extraction.
\citet{zhou2020fusion} use attention and tumor-region context
constraints, while Tri-Attention \citep{zhou2022triattention} weights
modality and spatial features and models correlations between sequences.

Transformer fusion models dependencies between modality
representations through attention.
NestedFormer \citep{xing2022nestedformer} uses intra-modality and
cross-modality attention at higher levels and modality-sensitive
gating at lower scales.
\citet{zhou2025deepfeatures} combine modality-wise, spatial, and
channel-wise feature fusion.
These designs adjust how features are combined for complete inputs.
With missing sequences, fusion must also accept different input subsets.

\subsection{Incomplete multimodal brain tumor segmentation}

Supporting different subsets first requires a representation that
can be constructed from the available inputs.
HeMIS \citep{havaei2016hemis} aggregates modality-specific features
through their mean and variance.
U-HVED \citep{dorent2019uhved} instead combines distributions inferred
from the available modalities in a probabilistic latent space
for joint segmentation and modality completion.
Beyond accepting different subsets, fusion can also adjust the
contribution of the remaining modalities.
RobustSeg \citep{chen2019robustseg} separates shared and
modality-specific representations and applies gated fusion.
RFNet \citep{ding2021rfnet} makes fusion region-aware, accounting
for differences in the information each sequence provides about
each tumor region.
D$^2$-Net \citep{yang2022d2net} models modality and region relationships
through dual disentanglement.

Training can provide additional information for incomplete inputs.
ACN \citep{wang2021acn} couples complete and incomplete configurations
through adversarial co-training and representation alignment.
LCKD \citep{wang2023lckd} selects teacher modalities according to
task performance, so supervision can differ between targets.
\citet{liu2025pretrainingkd} use complete-input predictions to guide
partial-input predictions during post-training.

Self-supervised learning (SSL) addresses feature learning before
supervised segmentation.
M3AE \citep{liu2023m3ae} reconstructs masked modalities and image patches
during pretraining.
It also uses model inversion for modality completion and
self-distillation during segmentation training.
M3AE combines pretraining with knowledge transfer.
UniME \citep{song2026unime} combines a shared ViT pretrained through
masked image modeling with multi-scale features from modality-specific
CNN encoders.
The pretrained representation provides global context, while
the CNN encoders supply features at higher spatial resolutions.

Within the network, interaction modules determine how features
exchange information.
mmFormer \citep{zhang2022mmformer} uses intra-modal and inter-modal
transformers, while M$^2$FTrans \citep{shi2024m2ftrans} incorporates
modality masking into transformer fusion to handle different
missing patterns.
Feature interaction is also modeled with state-space blocks.
LS$^3$M \citep{zhang2025ls3m} combines differentiable sequence reordering
with state-space blocks for local and long-range dependencies.
CDMS \citep{liu2025cdms} uses learnable consistency features and
fusion at multiple scales.
LS$^3$M focuses on spatial ordering, while CDMS incorporates
consistency information into fusion.

Fusion can also receive conditioning information.
MIFPN \citep{diao2025mifpn} derives prompts from modality-invariant
features.
MPDF-UNET \citep{xiao2025mpdfunet} combines learned modality priors
with input-dependent features and changes modality-combination
sampling during training.
MGML \citep{zou2026mgml} uses meta-parameterized fusion to construct
soft targets from available modalities, together with consistency
regularization.

TG-BWSA combines compressed modality features in a
shared adapter and returns adapted features to each encoder
stream \citep{rostamian2026tgbwsa}.
It learns two scalar weights per modality and stage to combine
the original and adapted features.
These scalar weights remain fixed across test inputs.
PMTM instead updates a tumor state from the available modalities
and uses that state to guide fusion.

Alongside these fusion mechanisms, we examine how an explicit tumor
state can accumulate evidence across encoder stages and guide fusion
for different available modality subsets.
This connects incomplete multimodal fusion to memory mechanisms
in segmentation.

\subsection{Memory mechanisms in medical image segmentation}

Memory provides context whose role depends on what is stored
and where it is reused.
Memory-Net \citep{wang2023memorynet} propagates cell memories through
hierarchically stacked gated convolutional units and across PET/CT
encoder branches.
HMDM-Net \citep{zhang2026hmdm} retains deep semantic representations
across decoder resolutions.
Memory-Net reuses intermediate information during feature extraction,
while HMDM-Net combines deep and shallow features during decoding.

Memory can also retain information across examples.
Sim-MPNet \citep{tang2025simmpnet} updates a prototype bank of
category features associated with lesions or organs during training.
The selective memory network \citep{zhao2026selectivememory} stores
and retrieves representative global features across samples
for 3D MRI segmentation.
Both provide category or global context beyond the features
of the current input.
In multimodal brain tumor segmentation, MSM-Seg
\citep{luo2025msmseg} organizes memory around modality and slice
relationships.
Its memory attention models correlations between modalities
and dependencies across slices.
The method also uses category-agnostic prompting and
a modality-adaptive fusion decoder.

PMTM builds an ordered tumor state from the available sequences
of the current subject, updates it across encoder stages, and
uses it to guide modality fusion at each stage.

\section{Method}
\label{sec:method}

In this work, we develop PMTM for brain tumor segmentation from
incomplete multimodal MRI. We first describe the overall network
architecture, followed by the progressive tumor-state memory and
state-guided fusion with modality-token interaction. We then present
the modality sampling procedure and training objectives.

\subsection{Network overview}

\begin{figure}[!htbp]
\centering
\includegraphics[width=0.95\textwidth]{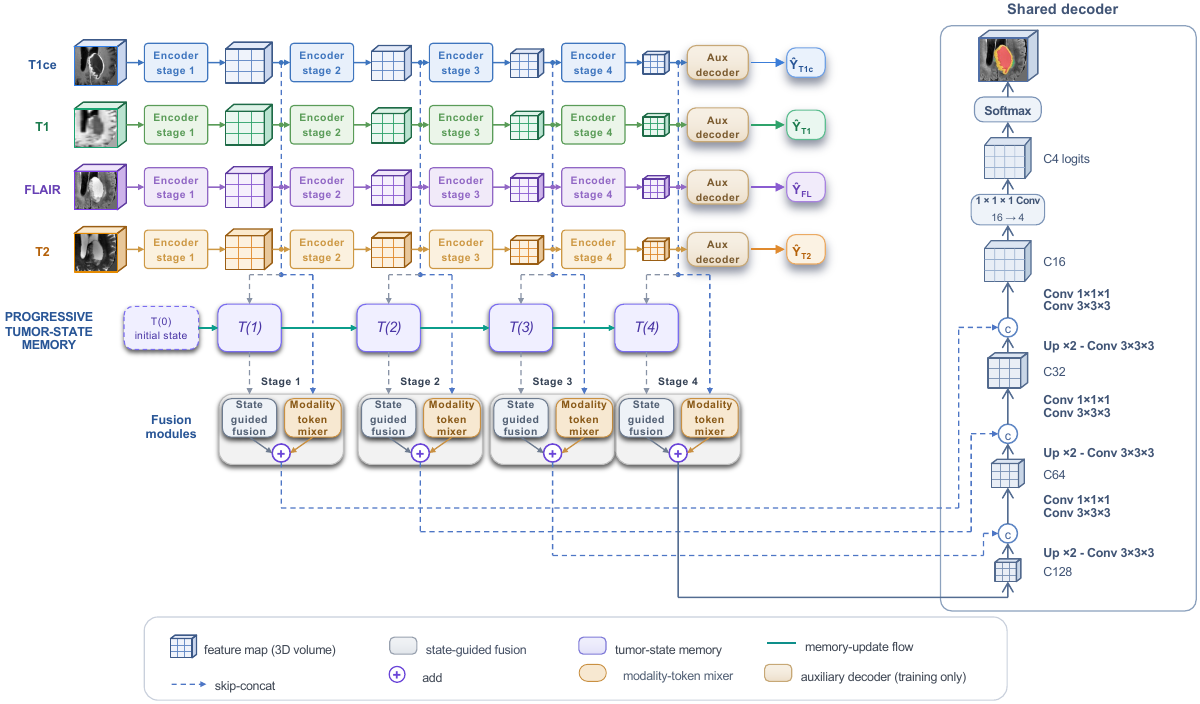}
\caption{Overview of PMTM. Four modality-specific encoders produce
multiscale features. At each stage, evidence from the available
modalities updates the tumor state, which guides fusion with
modality-token interaction. The fused features from all four stages
are passed to a shared decoder. Auxiliary decoders supervise each
modality branch during training and are discarded at inference.
The auxiliary skip connections and token-mixer internals
are omitted from the diagram.
}
\label{fig:pmsm_overview}
\end{figure}

Figure~\ref{fig:pmsm_overview} shows the overall architecture.
PMTM has four modality-specific encoders and a shared segmentation
decoder. The encoders have the same architecture but independent
parameters. Each encoder has four stages that produce feature maps
at different spatial scales.

For the current subject, PMTM maintains a tumor state that stores
evidence from the available MRI sequences. At each encoder stage,
the available modality features update this state.
Each stage has a fusion module with state-guided fusion and
a modality-token mixer. The updated state guides feature fusion,
and the mixer models interactions among the available modalities.
The shared decoder
combines the fused features from all four stages and predicts
four segmentation classes.

During training, each modality branch also has an auxiliary decoder
with the same architecture as the shared decoder but independent
parameters. It takes the deepest features of its own encoder as input
and uses the shallower scales through skip connections.
Its prediction supervises that modality branch directly.
The auxiliary decoders are discarded at inference.

Let
$\mathcal M=(\mathrm{T1ce},\mathrm{T1},\mathrm{FLAIR},\mathrm{T2})$
denote the ordered set of $M=4$ MRI modalities, and let
$S\subseteq\mathcal M$ denote the non-empty subset available for
a subject. Each modality-specific encoder $E_m$ produces features
$F_m^s\in\mathbb R^{C_s\times H_s\times W_s\times D_s}$
at stage $s\in\{1,2,3,4\}$. Only modalities in $S$ contribute
to state construction and feature fusion.

\subsection{Progressive tumor-state memory}
\label{sec:progressive_memory}

The tumor state stores evidence from the available MRI sequences.
The memory mechanism updates this state across encoder stages,
allowing fusion to use both the current features and evidence
retained from preceding stages.

\begin{figure}[!htbp]
\centering
\includegraphics[
    width=0.88\textwidth,
    height=0.64\textheight,
    keepaspectratio
]{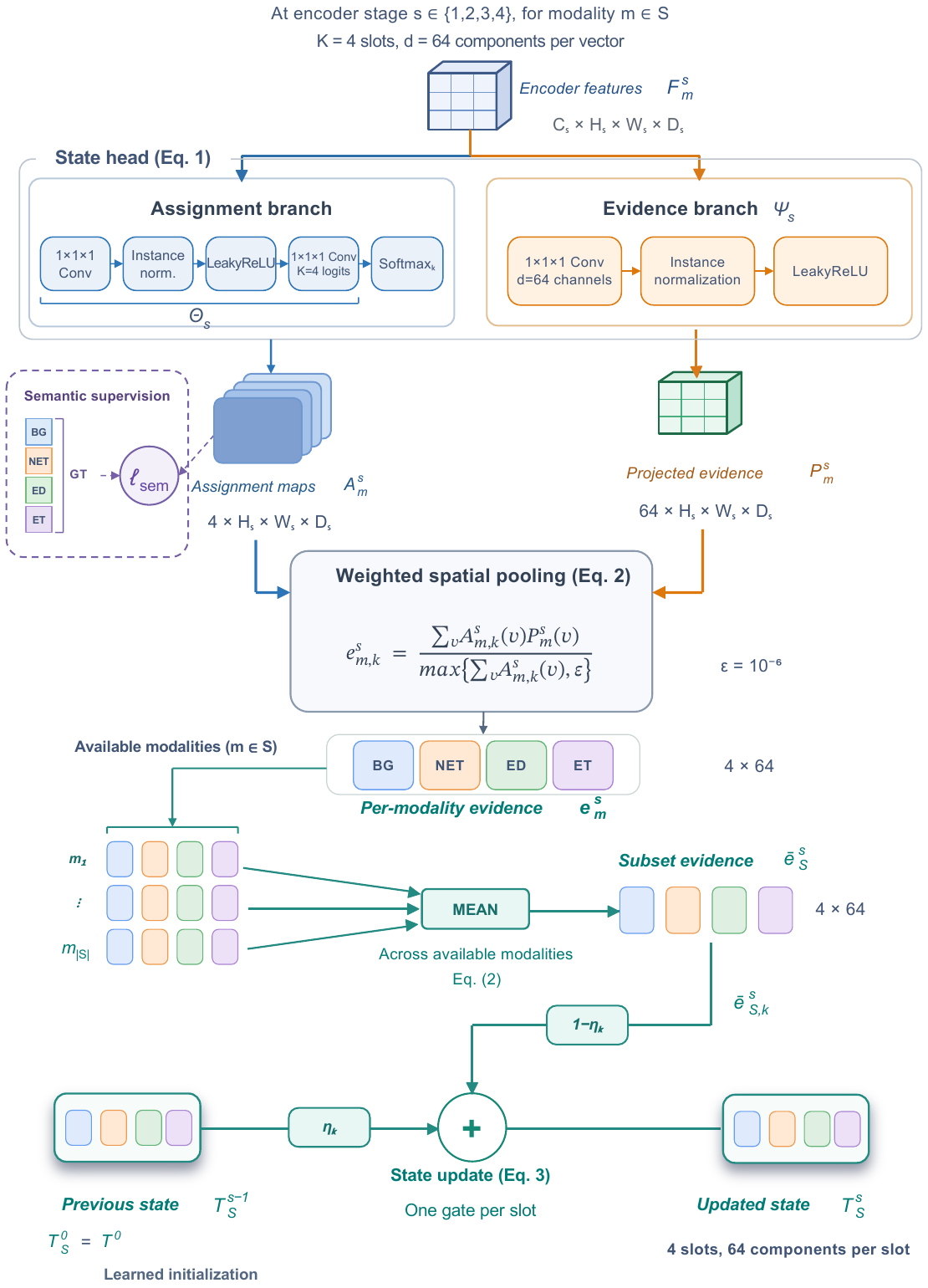}
\caption{Progressive tumor-state update at stage $s$. The state
head has an assignment branch and an evidence branch
(Eq.~\eqref{eq:state_head_outputs}). Weighted spatial pooling
produces four 64-dimensional vectors per modality. The MEAN block
averages corresponding vectors over $m\in S$
(Eq.~\eqref{eq:evidence_pool}). Slot-specific gates combine this
evidence with the previous state (Eq.~\eqref{eq:memory_update}).
}
\label{fig:memory_update}
\end{figure}

Figure~\ref{fig:memory_update} shows three operations in each update.
First, assignment maps pool each available modality's features
into an ordered set of evidence vectors.
We then average corresponding vectors across the available modalities.
Finally, a learned gate combines this evidence with the previous state.

\subsubsection{Semantic evidence construction}

The tumor state is an ordered set of $K=4$ feature vectors,
called state slots. Each vector has dimension $d=64$.
The assignment channels and state slots are intended to follow
the class order BG, NCR/NET, ED, and ET.
Here, BG is background, NCR/NET is the necrotic and non-enhancing
tumor core, ED is peritumoral edema, and ET is enhancing tumor.
We apply semantic supervision to the assignment maps as described
in Section~\ref{sec:training_objective}.

At stage $s$, a state head processes each available modality's
feature map. Its assignment branch produces a weight for each
slot at every spatial location. Its evidence branch projects
the features into the state space. The outputs are
\begin{equation}
\begin{aligned}
A_m^s
&=
\operatorname{Softmax}_{k}\!\left(\Theta_s(F_m^s)\right)
\in[0,1]^{K\times H_s\times W_s\times D_s},\\
P_m^s
&=
\Psi_s(F_m^s)
\in\mathbb R^{d\times H_s\times W_s\times D_s}.
\end{aligned}
\label{eq:state_head_outputs}
\end{equation}
Here, $A_m^s$ is the assignment map and $P_m^s$ is the projected
evidence tensor. The softmax operates over the $K$ assignment
channels, so their weights sum to one at each spatial location.

Within a stage, the state head is shared across modality branches.
Different stages have separate head parameters.
Both branches use $1\times1\times1$ convolutions.
The layer composition is given in Section~\ref{sec:experiments}.

We next use each assignment channel to pool the projected features
into one evidence vector. Let $v$ index the
$N_s=H_sW_sD_s$ spatial locations. For slot $k$, the evidence
within modality $m$ and its average across the available
modalities are
\begin{equation}
\begin{aligned}
e_{m,k}^s
&=
\frac{\sum_v A_{m,k}^s(v)P_m^s(v)}
{\max\{\sum_v A_{m,k}^s(v),\epsilon\}},
\qquad \epsilon=10^{-6},\\
\bar e_{S,k}^s
&=\frac{1}{|S|}\sum_{m\in S}e_{m,k}^s.
\end{aligned}
\label{eq:evidence_pool}
\end{equation}
Each assignment map therefore pools the projected features
of its own modality. Only the resulting evidence vectors
are averaged across the available modalities.

\subsubsection{Progressive state update}

We use these evidence vectors to update the previous state.
The initial state is a learned tensor
$T^0\in\mathbb R^{K\times d}$.
For each slot, a learned gate
$\eta_k=\sigma(\widehat\eta_k)$ weights the previous state
and the current evidence. Here, $\sigma$ denotes the sigmoid function.
With $T_{S,k}^{0}=T_k^0$, the update is
\begin{equation}
T_{S,k}^{s}
=
\eta_kT_{S,k}^{s-1}
+
(1-\eta_k)\bar e_{S,k}^{s},
\qquad
s\in\{1,2,3,4\}.
\label{eq:memory_update}
\end{equation}

A larger $\eta_k$ gives more weight to the previous state,
while a smaller value gives more weight to the current evidence.
Each slot has its own gate, shared across stages.
The first stage updates the learned initial state.
Each later stage updates the state produced by the preceding stage.

The updated state $T_S^s$ is used immediately by the fusion
module at stage $s$. At the beginning of every forward pass,
the memory is initialized from $T^0$.
The updated state is not carried between subjects.

\subsection{State-guided fusion}
\label{sec:state_fusion}

At each encoder stage, the fusion module combines the available
modality features through a base path and a gated residual path.
The base path averages projected features from the available
modalities. The residual path uses two channel gates to adjust
each modality's contribution.

\begin{figure}[!htbp]
\centering
\includegraphics[width=0.95\textwidth]{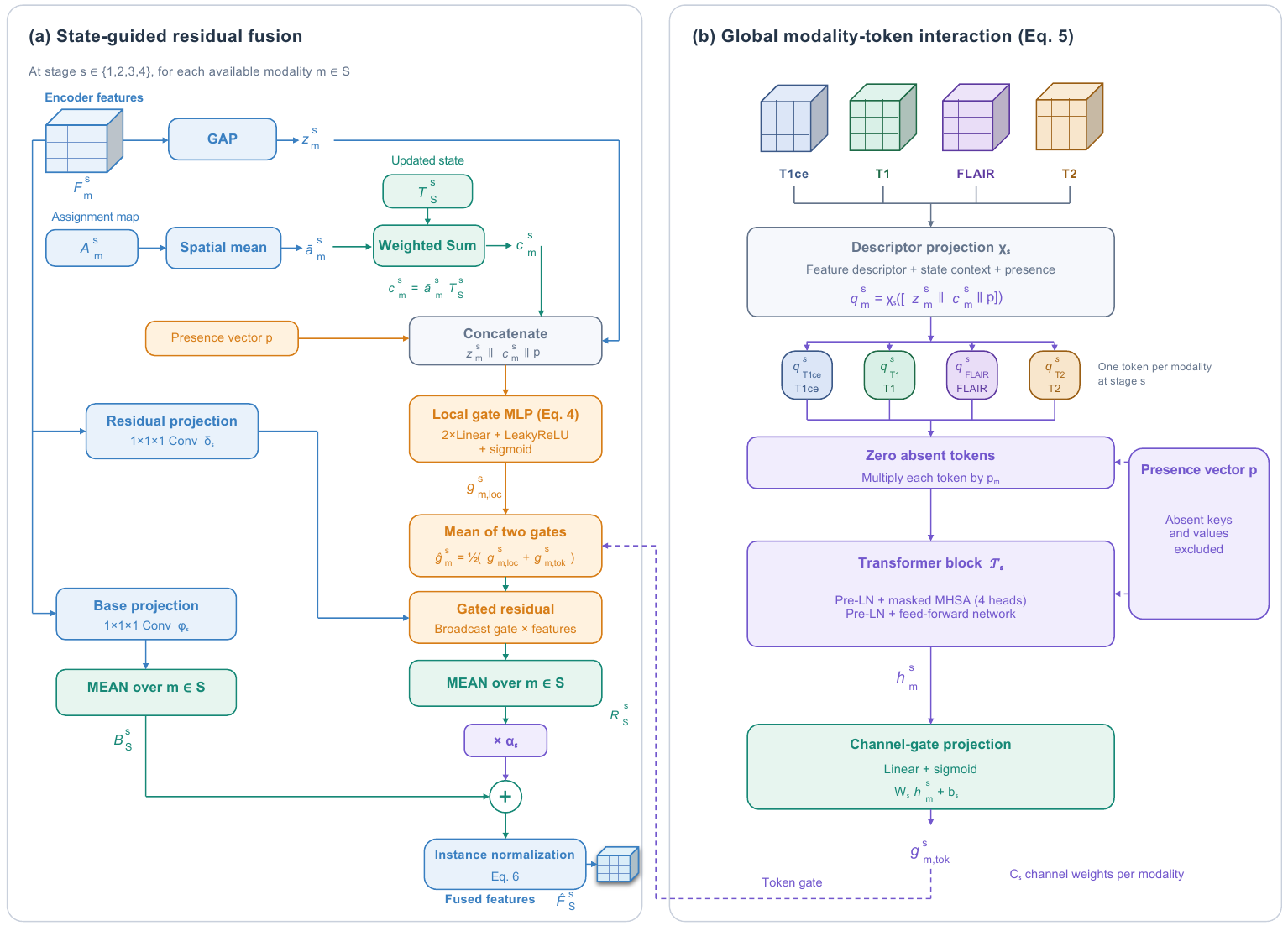}
\caption{State-guided fusion and modality-token interaction.
Panel (a) shows the base path, local gate, residual averaging,
scaling, and instance normalization
(Eqs.~\eqref{eq:state_guided_descriptors} and~\eqref{eq:residual_fusion}).
Panel (b) shows descriptor projection, masked global interaction
among modality tokens, and the token gates
(Eq.~\eqref{eq:token_mixer}).
Local and token gates are averaged before weighting residual features.
}
\label{fig:state_guided_fusion}
\end{figure}

The local gate uses a modality feature descriptor, a context
vector read from the updated tumor state, and the modality
availability pattern. The token gate uses interactions among
the available modality tokens. We average the two gates to
weight each modality's residual features. The weighted residuals
are then averaged over the available modalities, scaled, and added
to the base feature.
Figure~\ref{fig:state_guided_fusion} illustrates the local gate,
modality-token mixer, and residual path.

\subsubsection{State-guided residual fusion}

We encode modality availability with a binary vector
$\mathbf p\in\{0,1\}^{M}$, where $p_m=1$ when modality $m$
is available. The base feature $B_S^s$ averages the projected
features of the available modalities.

To compute the local gate, we first spatially pool $F_m^s$
into a feature descriptor $z_m^s$.
We also average the assignment map over spatial locations
to obtain slot weights $\bar a_m^s$.
These weights combine the updated state slots into a context
vector $c_m^s$.
We concatenate the feature descriptor, context vector,
and presence vector and pass them to an MLP.
These operations are written as
\begin{equation}
\begin{aligned}
B_S^s
&=\frac{1}{|S|}\sum_{m\in S}\phi_s(F_m^s),\\
z_m^s
&=\operatorname{GAP}(F_m^s),
&\bar a_m^s&=\frac{1}{N_s}\sum_vA_m^s(v),\\
c_m^s
&=\bar a_m^sT_S^s,
&g_{m,\mathrm{loc}}^s
&=\sigma\!\left(\operatorname{MLP}_s
[z_m^s\|c_m^s\|\mathbf p]\right).
\end{aligned}
\label{eq:state_guided_descriptors}
\end{equation}

Here, $\phi_s$ is a $1\times1\times1$ channel projection.
$\operatorname{GAP}$ denotes global average pooling,
and $\|$ denotes concatenation.
The two-layer MLP is shared across modality branches within a stage.
It produces $C_s$ channel logits.
The sigmoid converts these logits to the local gate
$g_{m,\mathrm{loc}}^s\in(0,1)^{C_s}$.

All modalities read from the same updated state $T_S^s$.
Their context vectors can differ because each modality uses
its own slot weights $\bar a_m^s$.

\subsubsection{Modality-token interaction}

The modality-token mixer allows the available modality tokens
to interact before residual fusion.
This interaction is global across the available modalities
within the same encoder stage.
At each stage, we construct one token per modality from
the feature descriptor, state context, and presence vector
used by the local gate.
A single transformer encoder block
\citep{vaswani2017attention} processes the tokens.
Its settings are given in Section~\ref{sec:experiments}.
A shared projection then maps each output to a channel gate.
The token and gate computations are
\begin{equation}
\begin{aligned}
q_m^s
&=\chi_s([z_m^s\|c_m^s\|\mathbf p]),\\
\{h_m^s\}_{m=1}^{M}
&=\mathcal T_s(\{q_m^s\}_{m=1}^{M};\mathbf p),\\
g_{m,\mathrm{tok}}^s
&=\sigma(W_sh_m^s+b_s).
\end{aligned}
\label{eq:token_mixer}
\end{equation}

Here, $\chi_s$ is a learned linear projection.
It maps the concatenated descriptor of dimension $C_s+d+M$
to a token $q_m^s\in\mathbb R^{C_s}$.
The transformer block $\mathcal T_s$ produces an updated token
$h_m^s\in\mathbb R^{C_s}$.
The learned parameters $W_s\in\mathbb R^{C_s\times C_s}$
and $b_s\in\mathbb R^{C_s}$ map each updated token to channel logits.
The sigmoid gives a gate
$g_{m,\mathrm{tok}}^s\in(0,1)^{C_s}$.
The subscript $\mathrm{tok}$ identifies the token gate;
$\mathrm{loc}$ in Eq.~\eqref{eq:state_guided_descriptors}
identifies the local gate.
Both projections are shared across modalities within a stage.

The mixer keeps $M$ token positions in the fixed modality order.
Before attention, each token is multiplied by its presence
indicator $p_m$.
A key-padding mask derived from $\mathbf p$ excludes absent
modalities from the attention keys and values.
Their output gates are also excluded from the residual average.
Only the available modalities contribute to the interaction
used for fusion.

\subsubsection{Fusion output}

We average the local and token gates to obtain the channel
weights for each modality's residual features. The gated
residuals are then averaged over the available modalities
and added to the base feature.
\begin{equation}
\begin{aligned}
\widehat g_m^s
&=\tfrac12(g_{m,\mathrm{loc}}^s+g_{m,\mathrm{tok}}^s),\\
R_S^s
&=\frac{1}{|S|}\sum_{m\in S}
\widehat g_m^s\odot\delta_s(F_m^s),\\
\widehat F_S^s
&=\operatorname{IN}_s(B_S^s+\alpha_sR_S^s),
\qquad \alpha_s=\sigma(\widehat\alpha_s).
\end{aligned}
\label{eq:residual_fusion}
\end{equation}
Here, $\delta_s$ is a $1\times1\times1$ projection with
parameters separate from the base projection $\phi_s$.
The symbol $\odot$ denotes multiplication along the channel dimension.
The channel weights are broadcast over spatial locations.
Each stage has its own learned residual scale $\alpha_s$.
We apply instance normalization after adding the scaled residual
to the base feature.
The shared decoder receives the fused features $\widehat F_S^s$
from all four stages.

\subsection{Training objective}
\label{sec:training_objective}

During training, we sample the available subset $S$ uniformly
for each example from the $2^M-1=15$ non-empty modality
combinations. Let
$Y\in\{0,1\}^{4\times H\times W\times D}$ denote the one-hot
segmentation target and $\widehat Y_S$ the class probabilities
predicted from subset $S$.

Training supervises the shared decoder prediction, the auxiliary
predictions from the available modalities, and the assignment maps
at all four stages.
The total objective is
\begin{equation}
\mathcal L
=
\mathcal L_{\mathrm{seg}}
+
\lambda_{\mathrm{aux}}(u)\mathcal L_{\mathrm{aux}}
+
\lambda_{\mathrm{state}}\mathcal L_{\mathrm{state}}.
\label{eq:full_objective}
\end{equation}
The three terms are the segmentation loss
$\mathcal L_{\mathrm{seg}}$, auxiliary loss
$\mathcal L_{\mathrm{aux}}$, and state loss
$\mathcal L_{\mathrm{state}}$.

Here, $u\in(0,1]$ is the fraction of completed training.
The objectives below are expressed for a single subject.
The loss coefficients and auxiliary schedule are given in
Section~\ref{sec:experiments}.

\subsubsection{Segmentation and auxiliary objectives}

The primary segmentation loss combines weighted cross-entropy,
multiclass soft Dice, and a region loss.
We define it as
$\mathcal L_{\mathrm{seg}}
=\mathcal L_{\mathrm{wce}}
+\mathcal L_{\mathrm{dice}}
+\lambda_{\mathrm{region}}\mathcal L_{\mathrm{region}}$.
These terms are evaluated using the shared decoder prediction
$\widehat Y_S$ and target $Y$.
The cross-entropy and multiclass Dice terms operate on the
four disjoint classes. For class $c$, the cross-entropy weight
is $w_c=1-N^{-1}\sum_vY_c(v)$, where $N=HWD$.

The region loss operates on the BraTS regions WT, TC, and ET.
WT combines NCR/NET, ED, and ET.
TC combines NCR/NET and ET, and ET is the enhancing class itself.
We use binary soft Dice for WT and TC.
ET can occupy a small part of the tumor, so we use
focal Tversky loss \citep{abraham2019focaltversky} for this region.
We set the Tversky weights to penalize missed enhancing tumor
more than false positives.
The parameter values are given in Section~\ref{sec:experiments}.

For a generic four-class probability map $\widehat Y$,
the region objective is
\begin{equation}
\begin{aligned}
\mathcal L_{\mathrm{region}}(\widehat Y,Y)
={}&
\omega_{\mathrm{WT}}\,
\ell_{\mathrm{BD}}(\widehat Y_{\mathrm{WT}},Y_{\mathrm{WT}})
\\ &+
\omega_{\mathrm{TC}}\,
\ell_{\mathrm{BD}}(\widehat Y_{\mathrm{TC}},Y_{\mathrm{TC}})
\\ &+
\omega_{\mathrm{ET}}\,
\ell_{\mathrm{FT}}(\widehat Y_{\mathrm{ET}},Y_{\mathrm{ET}}).
\end{aligned}
\label{eq:region_objective}
\end{equation}
Here, $\ell_{\mathrm{BD}}$ denotes binary soft Dice loss
and $\ell_{\mathrm{FT}}$ denotes focal Tversky loss.
Region probabilities and targets are obtained by summing
the corresponding disjoint-class channels.
The coefficients $\omega_{\mathrm{WT}}$,
$\omega_{\mathrm{TC}}$, and $\omega_{\mathrm{ET}}$ weight
the three region terms. Their values and the focal Tversky
parameters are reported in Section~\ref{sec:experiments}.

Let $\widehat Y_m^{\mathrm{aux}}$ denote the class probabilities
predicted by the auxiliary decoder for modality $m$.
The auxiliary objective averages the segmentation terms
over the available modalities.
\begin{equation}
\begin{aligned}
\mathcal L_{\mathrm{aux}}
=\frac{1}{|S|}\sum_{m\in S}\Big[
&\mathcal L_{\mathrm{wce}}(\widehat Y_m^{\mathrm{aux}},Y)
+\mathcal L_{\mathrm{dice}}(\widehat Y_m^{\mathrm{aux}},Y)
\\
&+\rho_{\mathrm{aux}}\lambda_{\mathrm{region}}\,
\mathcal L_{\mathrm{region}}(\widehat Y_m^{\mathrm{aux}},Y)
\Big].
\end{aligned}
\label{eq:auxiliary_objective}
\end{equation}
Here, $\rho_{\mathrm{aux}}$ scales the region contribution
relative to the primary segmentation objective.
The coefficient $\lambda_{\mathrm{aux}}(u)$ in
Eq.~\eqref{eq:full_objective} weights the entire auxiliary
objective during training.

\subsubsection{Semantic assignment}

The state loss gives semantic supervision to the assignment maps,
encouraging each channel to correspond to its intended class.
At each stage, we area-average the segmentation target to the
resolution of the assignment map.
This produces local class proportions $Y^s$.
We renormalize the target channels to correct small floating-point
deviations from a unit sum at each location.

For an assignment map $A$ at stage $s$, the semantic loss
for a single subject is
\begin{equation}
\begin{aligned}
\ell_{\mathrm{sem}}(A,Y^s)
={}&
-\frac{
\sum_{c,v}w_c^{\mathrm{sem}}Y_c^s(v)
\log\!\big(\max\{A_c(v),\epsilon\}\big)
}{
\max\{\sum_{c,v}w_c^{\mathrm{sem}}Y_c^s(v),\epsilon\}
}
\\ &+
\sum_c\widetilde w_c^{\mathrm{sem}}
\left(
1-
\frac{
2\sum_v A_c(v)Y_c^s(v)+\epsilon
}{
\sum_v A_c(v)+\sum_v Y_c^s(v)+\epsilon
}
\right),
\end{aligned}
\label{eq:semantic_map_loss}
\end{equation}
where $c$ indexes the four semantic classes, $v$ indexes
spatial locations at stage $s$, and $\epsilon=10^{-6}$.
The fixed class weights are $w_c^{\mathrm{sem}}$.
For the Dice term, we normalize them as
$\widetilde w_c^{\mathrm{sem}}
=w_c^{\mathrm{sem}}/\sum_j w_j^{\mathrm{sem}}$.
The first term is cross-entropy normalized by the weighted target mass.
The second term is class-weighted soft Dice loss.
The semantic class weights are reported in
Section~\ref{sec:experiments}.

We apply $\ell_{\mathrm{sem}}$ to the assignment map of
each available modality and to their mean,
$A_S^s=|S|^{-1}\sum_{m\in S}A_m^s$.
The resulting state objective is
\begin{equation}
\mathcal L_{\mathrm{state}}
=
\frac{1}{4}\sum_{s=1}^{4}
\frac{\ell_{\mathrm{sem}}(A_S^s,Y^s)
+\sum_{m\in S}\ell_{\mathrm{sem}}(A_m^s,Y^s)}{|S|+1}.
\label{eq:state_loss}
\end{equation}

Within each stage, the loss averages the $|S|$ modality
assignment terms and the term for their mean. The outer
average gives equal weight to all four encoder stages.

\FloatBarrier

\section{Experiments}
\label{sec:experiments}

We evaluate PMTM on BraTS2018, BraTS2020, and BraTS2023 GLI
across all 15 non-empty modality subsets. We first describe
the datasets, evaluation protocol, and implementation details.
We then compare PMTM with RobustSeg, RFNet, and mmFormer
trained in this study and with published BraTS2020 and BraTS2023
results~\citep{zou2026mgml,zhang2025ls3m}.
We also report component ablations, analyses of assignment maps
and tumor states, and segmentation maps under complete and
incomplete inputs.

\subsection{Datasets and evaluation protocol}

We use BraTS2018 and BraTS2020
\citep{menze2015brats,bakas2017advancing,bakas2018identifying}, and the adult glioma
task of the BraTS 2023 Challenge (BraTS2023 GLI)
\citep{brats2023gli,baid2021brats,baid2023dataset}.
The training/validation/test splits contain 205/23/57,
219/50/100, and 875/125/251 subjects, respectively.
For BraTS2020, we use the public subject lists from the official
M$^2$FTrans repository~\citep{shi2024m2ftrans}; for BraTS2023 GLI,
we use the public IM-Fuse subject lists~\citep{pipoli2025imfuse}.
Within each dataset, all methods trained in this study and component
ablations use the same fixed subject-level partitions.
Published results retain the training and evaluation protocols
reported in their source studies.

We evaluate Dice for WT, TC, and ET under each of the 15
non-empty subsets of T1ce, T1, FLAIR, and T2.
When the predicted and ground-truth regions are both empty,
Dice is set to one. When only one is empty, Dice is set to zero.
The overall score is Dice averaged equally over test subjects,
the three tumor regions, and all 15 modality subsets.
Dice scores are reported as percentages, and differences
between scores are reported in percentage points.

\subsection{Implementation details}

Each volume is cropped to a common foreground bounding box
defined by the nonzero voxels of the four MRI sequences.
The box is expanded to at least 128 voxels along each axis
when necessary.
Each sequence is standardized over its foreground voxels,
with background values kept at zero.

PMTM training uses random $128^3$ crops, rotations up to
$10^\circ$, intensity scaling for each modality in $[0.9,1.1]$,
and intensity shifts in $[-0.1,0.1]$.
Each spatial axis is independently flipped with probability 0.5.
Rotations use linear interpolation for MRI volumes and
nearest-neighbor interpolation for segmentation labels.

The four encoder stages have channel widths of 16, 32, 64,
and 128. Each stage contains three $3\times3\times3$
convolutional blocks with instance normalization and LeakyReLU.
The first block of stages~2--4 uses stride two.
The decoder uses trilinear upsampling and skip concatenation.
Its convolutional refinement layers have channel widths of
64, 32, and 16. A $1\times1\times1$ convolution maps the final
16 feature channels to four class logits. Softmax converts these
logits to voxel-wise class probabilities.
The auxiliary decoders use the same architecture,
as described in Section~\ref{sec:method}.

The assignment branch $\Theta_s$ uses a $1\times1\times1$
convolution, instance normalization, and LeakyReLU, followed
by a second $1\times1\times1$ convolution that produces $K$ logits.
The evidence branch $\Psi_s$ uses a $1\times1\times1$ convolution
followed by instance normalization and LeakyReLU to produce
$d=64$ evidence channels.
The modality-token mixer uses a single transformer encoder block
with four attention heads, a feed-forward width of
$\max(2C_s,64)$, GELU activation, pre-layer normalization,
and zero dropout.
Layer normalization is applied before the attention and
feed-forward sublayers (Pre-LN).
All LeakyReLU activations have a negative slope of 0.2.

PMTM is implemented in PyTorch and trained using Adam
with AMSGrad. The initial learning rate is $10^{-4}$
and follows polynomial decay with power 0.9.
Weight decay is $10^{-4}$, and the effective batch size is two.
PMTM experiments use one NVIDIA RTX A6000 GPU with 48\,GB memory.

In the experiments conducted in this study, all four methods
are trained for 1000 epochs on BraTS2018 and BraTS2020 and
for 200 epochs on BraTS2023 GLI.
We select the checkpoint with the highest average validation
Dice across all 15 modality subsets.
The comparisons use method-specific crop sizes of $80^3$
for RobustSeg and RFNet and $128^3$ for mmFormer and PMTM.
The sizes for the comparison methods follow their original
input configurations.

For the objective in Section~\ref{sec:training_objective},
we use $\lambda_{\mathrm{region}}=0.35$ and
$\lambda_{\mathrm{state}}=0.05$.
The region weights
$(\omega_{\mathrm{WT}},\omega_{\mathrm{TC}},\omega_{\mathrm{ET}})$
are $(0.20,0.30,0.50)$.
Focal Tversky loss uses
$\alpha_{\mathrm{FT}}=0.3$ for false positives,
$\beta_{\mathrm{FT}}=0.7$ for false negatives, and
$\gamma_{\mathrm{FT}}=0.75$.
The loss is $(1-\mathrm{TI})^{\gamma_{\mathrm{FT}}}$,
where $\mathrm{TI}$ is the soft Tversky index.

The auxiliary region multiplier is $\rho_{\mathrm{aux}}=0.15$,
so $\rho_{\mathrm{aux}}\lambda_{\mathrm{region}}=0.0525$.
The semantic class weights are $(0.25,1,1,1)$
for BG, NCR/NET, ED, and ET.
The auxiliary coefficient is 0.5 during the first 30\%
of training. It decreases linearly to 0.2 by 70\%
and then decreases linearly to 0.1 at the final epoch.

PMTM inference uses $128^3$ sliding windows with 50\% overlap.
The tumor state is initialized from $T^0$ separately
for each sliding window.
Class probabilities are averaged in overlapping regions
before applying argmax at each voxel.
No test-time augmentation is applied.

\subsection{Comparison with existing methods}
\label{sec:results}

We compare PMTM with RobustSeg~\citep{chen2019robustseg},
RFNet~\citep{ding2021rfnet}, and mmFormer~\citep{zhang2022mmformer}.
Table~\ref{tab:main_benchmark_all} reports the results averaged
over all 15 modality subsets on the three datasets.
PMTM obtains the highest overall average Dice among the
four methods on each dataset.
Among the baselines, mmFormer has the highest overall average.
The regional differences relative to mmFormer are larger for
TC and ET than for WT.
We additionally report published results for BraTS2020 and
BraTS2023 GLI in Tables~\ref{tab:brats2020_published_comparison}
and~\ref{tab:brats2023_published_comparison}.
The source column distinguishes our experiments from literature
results. Published regional means are preserved, and their
three-region averages are computed from those reported means.
Published results use the training and evaluation settings
reported in their original studies.

\begin{table*}[!htbp]
\centering
\caption{Results on the three BraTS cohorts, averaged over all 15 non-empty modality subsets. Dice values are percentages; PMTM gains are percentage-point differences. Crop denotes the side length, in voxels, of the cubic training input. Bold marks the highest score and underlining the second-highest distinct score within each dataset. Ties receive the same marking.}
\label{tab:main_benchmark_all}
\resizebox{\textwidth}{!}{%
\begin{tabular}{llccccccc}
\toprule
Dataset & Method & Epochs & Crop & WT & TC & ET & Avg. & PMTM gain \\
\midrule
\multirow{4}{*}{BraTS2018}
 & RobustSeg & 1000 & 80 & 84.55 & 69.11 & 51.28 & 68.32 & +6.99 \\
 & RFNet & 1000 & 80 & 85.68 & 74.86 & 57.17 & 72.57 & +2.74 \\
 & mmFormer & 1000 & 128 & \textbf{87.50} & \underline{74.92} & \underline{57.77} & \underline{73.40} & +1.91 \\
 & \textbf{Ours} & 1000 & 128 & \underline{86.79} & \textbf{77.17} & \textbf{61.95} & \textbf{75.31} & -- \\
\midrule
\multirow{4}{*}{BraTS2020}
 & RobustSeg & 1000 & 80 & 84.34 & 73.60 & 55.60 & 71.18 & +6.36 \\
 & RFNet & 1000 & 80 & 86.80 & 78.15 & 57.44 & 74.13 & +3.41 \\
 & mmFormer & 1000 & 128 & \underline{87.60} & \underline{79.32} & \underline{58.27} & \underline{75.06} & +2.48 \\
 & \textbf{Ours} & 1000 & 128 & \textbf{87.63} & \textbf{81.53} & \textbf{63.47} & \textbf{77.54} & -- \\
\midrule
\multirow{4}{*}{BraTS2023 GLI}
 & RobustSeg & 200 & 80 & 84.69 & 73.10 & 60.69 & 72.82 & +7.65 \\
 & RFNet & 200 & 80 & 85.68 & 75.66 & 61.71 & 74.35 & +6.12 \\
 & mmFormer & 200 & 128 & \underline{87.45} & \underline{79.42} & \underline{66.21} & \underline{77.69} & +2.78 \\
 & \textbf{Ours} & 200 & 128 & \textbf{87.74} & \textbf{82.72} & \textbf{70.96} & \textbf{80.47} & -- \\
\bottomrule
\end{tabular}%
}
\end{table*}

\subsubsection{BraTS2018}

On BraTS2018, PMTM obtains an average Dice of 75.31\%,
1.91 percentage points higher than mmFormer.
TC Dice is higher by 2.25 points and ET Dice by 4.18 points.
WT Dice is lower by 0.71 points.

Table~\ref{tab:brats2018_rfnet_style_missing_methods}
reports the results for individual modality subsets.
Among the four methods, PMTM has the highest three-region
average (Avg.) in all 15 subsets.
It has the highest ET Dice in all 15 subsets and the highest
TC Dice in 14 subsets.
mmFormer has the highest WT Dice in all 15 subsets
and the highest TC Dice for FLAIR-only input.

\newgeometry{left=24pt,right=24pt,top=50pt,bottom=50pt}
\begin{landscape}
\begin{table}[p]
\centering
\captionsetup{font=small}
\caption{BraTS2018 Dice (\%) for all 15 non-empty modality subsets.}
\label{tab:brats2018_rfnet_style_missing_methods}
\setlength{\arrayrulewidth}{0.25pt}
\setlength{\tabcolsep}{0.9pt}
\renewcommand{\arraystretch}{1.0}
\fontsize{8}{8.4}\selectfont
\makebox[\linewidth][c]{\resizebox{0.96\linewidth}{!}{%
\begin{tabular}{l|l|r|r|r|r|r|r|r|r|r|r|r|r|r|r|r|r}
\hline
\multirow{4}{*}{M} & \multicolumn{1}{c|}{FLAIR} & \multicolumn{1}{c|}{\ModPresent} & \multicolumn{1}{c|}{\ModAbsent} & \multicolumn{1}{c|}{\ModPresent} & \multicolumn{1}{c|}{\ModPresent} & \multicolumn{1}{c|}{\ModPresent} & \multicolumn{1}{c|}{\ModPresent} & \multicolumn{1}{c|}{\ModPresent} & \multicolumn{1}{c|}{\ModAbsent} & \multicolumn{1}{c|}{\ModPresent} & \multicolumn{1}{c|}{\ModAbsent} & \multicolumn{1}{c|}{\ModAbsent} & \multicolumn{1}{c|}{\ModPresent} & \multicolumn{1}{c|}{\ModAbsent} & \multicolumn{1}{c|}{\ModAbsent} & \multicolumn{1}{c|}{\ModAbsent} & \multirow{4}{*}{Mean} \\
 & \multicolumn{1}{c|}{T1ce} & \multicolumn{1}{c|}{\ModPresent} & \multicolumn{1}{c|}{\ModPresent} & \multicolumn{1}{c|}{\ModPresent} & \multicolumn{1}{c|}{\ModAbsent} & \multicolumn{1}{c|}{\ModPresent} & \multicolumn{1}{c|}{\ModPresent} & \multicolumn{1}{c|}{\ModAbsent} & \multicolumn{1}{c|}{\ModAbsent} & \multicolumn{1}{c|}{\ModAbsent} & \multicolumn{1}{c|}{\ModPresent} & \multicolumn{1}{c|}{\ModPresent} & \multicolumn{1}{c|}{\ModAbsent} & \multicolumn{1}{c|}{\ModAbsent} & \multicolumn{1}{c|}{\ModPresent} & \multicolumn{1}{c|}{\ModAbsent} &  \\
 & \multicolumn{1}{c|}{T1} & \multicolumn{1}{c|}{\ModPresent} & \multicolumn{1}{c|}{\ModPresent} & \multicolumn{1}{c|}{\ModAbsent} & \multicolumn{1}{c|}{\ModPresent} & \multicolumn{1}{c|}{\ModPresent} & \multicolumn{1}{c|}{\ModAbsent} & \multicolumn{1}{c|}{\ModAbsent} & \multicolumn{1}{c|}{\ModPresent} & \multicolumn{1}{c|}{\ModPresent} & \multicolumn{1}{c|}{\ModPresent} & \multicolumn{1}{c|}{\ModAbsent} & \multicolumn{1}{c|}{\ModAbsent} & \multicolumn{1}{c|}{\ModPresent} & \multicolumn{1}{c|}{\ModAbsent} & \multicolumn{1}{c|}{\ModAbsent} &  \\
 & \multicolumn{1}{c|}{T2} & \multicolumn{1}{c|}{\ModPresent} & \multicolumn{1}{c|}{\ModPresent} & \multicolumn{1}{c|}{\ModPresent} & \multicolumn{1}{c|}{\ModPresent} & \multicolumn{1}{c|}{\ModAbsent} & \multicolumn{1}{c|}{\ModAbsent} & \multicolumn{1}{c|}{\ModPresent} & \multicolumn{1}{c|}{\ModPresent} & \multicolumn{1}{c|}{\ModAbsent} & \multicolumn{1}{c|}{\ModAbsent} & \multicolumn{1}{c|}{\ModPresent} & \multicolumn{1}{c|}{\ModAbsent} & \multicolumn{1}{c|}{\ModAbsent} & \multicolumn{1}{c|}{\ModAbsent} & \multicolumn{1}{c|}{\ModPresent} &  \\
\hline
\multirow{4}{*}{WT} & RobustSeg~\citep{chen2019robustseg} & 89.63 & 86.18 & 89.39 & 88.99 & 88.91 & 88.69 & 88.46 & 84.95 & 88.42 & 77.33 & 85.36 & 85.86 & 70.25 & 73.46 & 82.40 & 84.55 \\
 & RFNet~\citep{ding2021rfnet} & 89.77 & 87.45 & 89.84 & 89.14 & 88.95 & 89.01 & 88.69 & 86.69 & 87.91 & 79.63 & 86.42 & 86.82 & 74.05 & 75.93 & 84.83 & 85.68 \\
 & mmFormer~\citep{zhang2022mmformer} & \textbf{90.95} & \textbf{88.47} & \textbf{90.95} & \textbf{90.84} & \textbf{90.39} & \textbf{90.48} & \textbf{90.49} & \textbf{88.16} & \textbf{90.26} & \textbf{80.88} & \textbf{88.16} & \textbf{89.42} & \textbf{78.08} & \textbf{78.19} & \textbf{86.73} & \textbf{87.50} \\
 & \textbf{Ours} & \underline{90.66} & \underline{87.49} & \underline{90.64} & \underline{90.30} & \underline{90.24} & \underline{90.16} & \underline{90.07} & \underline{87.11} & \underline{89.92} & \underline{79.87} & \underline{87.21} & \underline{88.82} & \underline{76.42} & \underline{77.08} & \underline{85.89} & \underline{86.79} \\
\hline
\multirow{4}{*}{TC} & RobustSeg~\citep{chen2019robustseg} & 81.02 & 80.49 & 80.88 & 64.51 & 81.22 & 68.68 & 61.28 & 62.31 & 60.80 & 78.88 & 80.36 & 53.68 & 48.00 & 76.98 & 57.60 & 69.11 \\
 & RFNet~\citep{ding2021rfnet} & \underline{82.09} & \underline{82.56} & 81.55 & \underline{70.97} & 81.35 & 80.54 & \underline{70.20} & \underline{69.77} & 68.80 & \underline{80.83} & \underline{81.74} & 65.03 & 60.45 & 79.70 & \underline{67.32} & 74.86 \\
 & mmFormer~\citep{zhang2022mmformer} & 81.23 & 81.37 & \underline{81.72} & 70.10 & \underline{81.59} & \underline{81.91} & 69.40 & 68.24 & \underline{70.31} & 79.95 & 81.62 & \textbf{66.46} & \underline{63.25} & \underline{79.96} & 66.75 & \underline{74.92} \\
 & \textbf{Ours} & \textbf{83.29} & \textbf{83.80} & \textbf{83.16} & \textbf{73.79} & \textbf{83.79} & \textbf{83.46} & \textbf{72.36} & \textbf{72.17} & \textbf{71.54} & \textbf{83.23} & \textbf{83.60} & \underline{66.27} & \textbf{63.43} & \textbf{82.86} & \textbf{70.87} & \textbf{77.17} \\
\hline
\multirow{4}{*}{ET} & RobustSeg~\citep{chen2019robustseg} & 71.84 & 71.24 & 71.02 & 36.48 & 70.92 & 70.44 & 34.67 & 32.07 & 32.19 & 69.20 & 69.85 & 25.74 & 17.32 & 67.20 & 29.03 & 51.28 \\
 & RFNet~\citep{ding2021rfnet} & 72.17 & 72.94 & 72.21 & \underline{45.08} & 71.77 & 71.84 & \underline{44.70} & \underline{43.75} & 38.22 & 71.28 & 72.03 & \underline{35.21} & 33.09 & 70.87 & \underline{42.33} & 57.17 \\
 & mmFormer~\citep{zhang2022mmformer} & \underline{72.84} & \underline{73.59} & \underline{73.10} & 43.89 & \underline{73.39} & \underline{73.50} & 43.00 & 43.44 & \underline{40.00} & \underline{73.02} & \underline{73.71} & 32.94 & \underline{36.99} & \underline{72.70} & 40.45 & \underline{57.77} \\
 & \textbf{Ours} & \textbf{74.83} & \textbf{74.66} & \textbf{74.83} & \textbf{52.49} & \textbf{74.98} & \textbf{74.82} & \textbf{50.94} & \textbf{50.71} & \textbf{47.89} & \textbf{74.38} & \textbf{74.57} & \textbf{42.54} & \textbf{39.11} & \textbf{73.92} & \textbf{48.61} & \textbf{61.95} \\
\hline
\multirow{4}{*}{Avg.} & RobustSeg~\citep{chen2019robustseg} & 80.83 & 79.30 & 80.43 & 63.33 & 80.35 & 75.94 & 61.47 & 59.78 & 60.47 & 75.14 & 78.52 & 55.09 & 45.19 & 72.55 & 56.34 & 68.32 \\
 & RFNet~\citep{ding2021rfnet} & 81.34 & 80.98 & 81.20 & \underline{68.40} & 80.69 & 80.46 & \underline{67.86} & \underline{66.74} & 64.98 & 77.25 & 80.06 & 62.35 & 55.86 & 75.50 & \underline{64.83} & 72.57 \\
 & mmFormer~\citep{zhang2022mmformer} & \underline{81.67} & \underline{81.14} & \underline{81.92} & 68.28 & \underline{81.79} & \underline{81.96} & 67.63 & 66.61 & \underline{66.86} & \underline{77.95} & \underline{81.16} & \underline{62.94} & \underline{59.44} & \underline{76.95} & 64.64 & \underline{73.40} \\
 & \textbf{Ours} & \textbf{82.93} & \textbf{81.98} & \textbf{82.88} & \textbf{72.19} & \textbf{83.00} & \textbf{82.82} & \textbf{71.12} & \textbf{69.99} & \textbf{69.79} & \textbf{79.16} & \textbf{81.79} & \textbf{65.87} & \textbf{59.65} & \textbf{77.96} & \textbf{68.46} & \textbf{75.31} \\
\hline
\end{tabular}%
}}
\par\vspace{1pt}
\begin{minipage}{\linewidth}
\fontsize{8}{9}\selectfont
\ModPresent\ = available modality; \ModAbsent\ = missing modality. M denotes the tumor-region metric. All results are from this study. References beside method names identify the original methods. Bold marks the highest score and underlining the second-highest distinct score among all methods shown. Ties receive the same marking. Avg. is the arithmetic mean of WT, TC, and ET for each subset. Mean denotes the average over all 15 subsets.
\end{minipage}
\end{table}
\end{landscape}
\restoregeometry
\FloatBarrier

\subsubsection{BraTS2020}

On BraTS2020, PMTM obtains an average Dice of 77.54\%,
2.48 percentage points higher than mmFormer.
WT Dice is higher by 0.03 points, TC Dice by 2.21 points,
and ET Dice by 5.20 points in our experiments.

Table~\ref{tab:brats2020_published_comparison} adds published
results from Zou et~al.~\citep{zou2026mgml} and
Zhang et~al.~\citep{zhang2025ls3m}.
PMTM has an average Dice of 77.54\%, compared with 77.19\%
for LS$^3$M and 76.53\% for IM-Fuse + MGML in these sources.
LS$^3$M has higher WT and ET Dice, while PMTM has higher TC Dice.

\begin{table}[!htbp]
\centering
\small
\setlength{\tabcolsep}{4pt}
\caption{BraTS2020 Dice (\%) averaged over all 15 non-empty modality subsets. The Source column identifies experiments conducted in this study and published results. Bold marks the highest score and underlining the second-highest distinct score among all methods shown. Ties receive the same marking.}
\label{tab:brats2020_published_comparison}
\begin{tabular}{llrrrr}
\toprule
Method & Source & WT $\uparrow$ & TC $\uparrow$ & ET $\uparrow$ & Avg. $\uparrow$ \\
\midrule
MMANet \citep{wei2023mmanet} & \citep{zou2026mgml} & 82.64 & 76.99 & 56.39 & 72.01 \\
PASSION \citep{shi2024passion} & \citep{zou2026mgml} & 84.28 & 71.37 & 53.92 & 69.86 \\
DMAF-Net \citep{lan2025dmaf} & \citep{zou2026mgml} & 84.71 & 71.98 & 54.30 & 70.33 \\
IM-Fuse \citep{pipoli2025imfuse} & \citep{zou2026mgml} & 87.08 & 79.01 & 61.70 & 75.93 \\
IM-Fuse + MGML \citep{zou2026mgml} & \citep{zou2026mgml} & 87.55 & 79.36 & 62.67 & 76.53 \\
LS$^3$M \citep{zhang2025ls3m} & \citep{zhang2025ls3m} & \textbf{88.22} & \underline{79.76} & \textbf{63.60} & \underline{77.19} \\
\midrule
RobustSeg \citep{chen2019robustseg} & This study & 84.34 & 73.60 & 55.60 & 71.18 \\
RFNet \citep{ding2021rfnet} & This study & 86.80 & 78.15 & 57.44 & 74.13 \\
mmFormer \citep{zhang2022mmformer} & This study & 87.60 & 79.32 & 58.27 & 75.06 \\
\textbf{Ours} & This study & \underline{87.63} & \textbf{81.53} & \underline{63.47} & \textbf{77.54} \\
\bottomrule
\end{tabular}
\end{table}

Table~\ref{tab:rfnet_style_missing_methods}
reports WT, TC, ET, and three-region averages for individual
modality subsets, including the published results.
Published WT, TC, and ET means are reproduced from the cited sources.

\newgeometry{left=24pt,right=24pt,top=50pt,bottom=50pt}
\begin{landscape}
\begin{table}[p]
\centering
\captionsetup{font=small}
\caption{BraTS2020 Dice (\%) for all 15 non-empty modality subsets.}
\label{tab:rfnet_style_missing_methods}
\setlength{\arrayrulewidth}{0.25pt}
\setlength{\tabcolsep}{0.9pt}
\renewcommand{\arraystretch}{1.0}
\fontsize{8}{8.4}\selectfont
\makebox[\linewidth][c]{\resizebox{0.96\linewidth}{!}{%
\begin{tabular}{l|l|r|r|r|r|r|r|r|r|r|r|r|r|r|r|r|r}
\hline
\multirow{4}{*}{M} & \multicolumn{1}{c|}{FLAIR} & \multicolumn{1}{c|}{\ModPresent} & \multicolumn{1}{c|}{\ModAbsent} & \multicolumn{1}{c|}{\ModPresent} & \multicolumn{1}{c|}{\ModPresent} & \multicolumn{1}{c|}{\ModPresent} & \multicolumn{1}{c|}{\ModPresent} & \multicolumn{1}{c|}{\ModPresent} & \multicolumn{1}{c|}{\ModAbsent} & \multicolumn{1}{c|}{\ModPresent} & \multicolumn{1}{c|}{\ModAbsent} & \multicolumn{1}{c|}{\ModAbsent} & \multicolumn{1}{c|}{\ModPresent} & \multicolumn{1}{c|}{\ModAbsent} & \multicolumn{1}{c|}{\ModAbsent} & \multicolumn{1}{c|}{\ModAbsent} & \multirow{4}{*}{Mean} \\
 & \multicolumn{1}{c|}{T1ce} & \multicolumn{1}{c|}{\ModPresent} & \multicolumn{1}{c|}{\ModPresent} & \multicolumn{1}{c|}{\ModPresent} & \multicolumn{1}{c|}{\ModAbsent} & \multicolumn{1}{c|}{\ModPresent} & \multicolumn{1}{c|}{\ModPresent} & \multicolumn{1}{c|}{\ModAbsent} & \multicolumn{1}{c|}{\ModAbsent} & \multicolumn{1}{c|}{\ModAbsent} & \multicolumn{1}{c|}{\ModPresent} & \multicolumn{1}{c|}{\ModPresent} & \multicolumn{1}{c|}{\ModAbsent} & \multicolumn{1}{c|}{\ModAbsent} & \multicolumn{1}{c|}{\ModPresent} & \multicolumn{1}{c|}{\ModAbsent} &  \\
 & \multicolumn{1}{c|}{T1} & \multicolumn{1}{c|}{\ModPresent} & \multicolumn{1}{c|}{\ModPresent} & \multicolumn{1}{c|}{\ModAbsent} & \multicolumn{1}{c|}{\ModPresent} & \multicolumn{1}{c|}{\ModPresent} & \multicolumn{1}{c|}{\ModAbsent} & \multicolumn{1}{c|}{\ModAbsent} & \multicolumn{1}{c|}{\ModPresent} & \multicolumn{1}{c|}{\ModPresent} & \multicolumn{1}{c|}{\ModPresent} & \multicolumn{1}{c|}{\ModAbsent} & \multicolumn{1}{c|}{\ModAbsent} & \multicolumn{1}{c|}{\ModPresent} & \multicolumn{1}{c|}{\ModAbsent} & \multicolumn{1}{c|}{\ModAbsent} &  \\
 & \multicolumn{1}{c|}{T2} & \multicolumn{1}{c|}{\ModPresent} & \multicolumn{1}{c|}{\ModPresent} & \multicolumn{1}{c|}{\ModPresent} & \multicolumn{1}{c|}{\ModPresent} & \multicolumn{1}{c|}{\ModAbsent} & \multicolumn{1}{c|}{\ModAbsent} & \multicolumn{1}{c|}{\ModPresent} & \multicolumn{1}{c|}{\ModPresent} & \multicolumn{1}{c|}{\ModAbsent} & \multicolumn{1}{c|}{\ModAbsent} & \multicolumn{1}{c|}{\ModPresent} & \multicolumn{1}{c|}{\ModAbsent} & \multicolumn{1}{c|}{\ModAbsent} & \multicolumn{1}{c|}{\ModAbsent} & \multicolumn{1}{c|}{\ModPresent} &  \\
\hline
\multirow{10}{*}{WT} & MMANet~\citep{wei2023mmanet}\textsuperscript{a} & 89.25 & 86.38 & 89.37 & 87.97 & 87.92 & 86.74 & 87.47 & 85.33 & 87.64 & 70.98 & 85.64 & 83.21 & 64.35 & 64.46 & 82.88 & 82.64 \\
 & PASSION~\citep{shi2024passion}\textsuperscript{a} & 89.74 & 89.36 & 88.00 & 88.82 & 88.65 & 76.30 & 85.62 & 87.83 & 87.71 & \textbf{86.90} & 86.26 & 70.51 & \underline{84.22} & 70.69 & 83.58 & 84.28 \\
 & DMAF-Net~\citep{lan2025dmaf}\textsuperscript{a} & 89.60 & \textbf{89.91} & 88.27 & 89.53 & 88.30 & 77.57 & 85.86 & \underline{88.25} & 87.60 & \textbf{86.90} & 86.67 & 72.92 & \textbf{84.94} & 70.64 & 83.64 & 84.71 \\
 & IM-Fuse~\citep{pipoli2025imfuse}\textsuperscript{a} & 91.02 & 88.79 & 90.44 & 90.73 & 90.44 & 89.57 & 89.90 & 88.03 & 89.81 & 81.28 & 88.01 & 87.55 & 77.52 & 76.90 & 86.29 & 87.08 \\
 & IM-Fuse + MGML~\citep{zou2026mgml}\textsuperscript{a} & 91.21 & 89.12 & 90.61 & \underline{90.98} & 90.73 & 89.95 & 90.11 & 88.14 & 89.93 & 82.26 & \underline{88.46} & 88.22 & 78.74 & 78.08 & \underline{86.80} & 87.55 \\
 & LS$^3$M~\citep{zhang2025ls3m}\textsuperscript{b} & \textbf{92.15} & \underline{89.75} & \textbf{91.87} & \textbf{91.45} & \textbf{91.52} & \textbf{91.00} & \textbf{91.45} & \textbf{88.88} & \textbf{90.57} & \underline{82.58} & \textbf{89.18} & \underline{88.66} & 79.32 & 78.24 & \textbf{87.29} & \textbf{88.22} \\
\cline{2-18}
 & RobustSeg~\citep{chen2019robustseg} & 89.65 & 86.80 & 88.86 & 89.42 & 89.05 & 87.50 & 88.27 & 85.70 & 88.28 & 76.99 & 86.14 & 83.04 & 71.55 & 71.53 & 82.36 & 84.34 \\
 & RFNet~\citep{ding2021rfnet} & 90.81 & 87.95 & 90.70 & 90.23 & 90.29 & 89.89 & 89.81 & 87.17 & 89.02 & 81.21 & 87.45 & 86.97 & 77.51 & 78.82 & 84.16 & 86.80 \\
 & mmFormer~\citep{zhang2022mmformer} & 91.15 & 88.50 & 91.01 & 90.52 & 90.82 & 90.58 & 90.06 & 87.43 & 89.74 & 82.16 & 88.00 & 88.41 & 79.80 & \textbf{80.00} & 85.85 & 87.60 \\
 & \textbf{Ours} & \underline{91.33} & 88.41 & \underline{91.16} & 90.78 & \underline{91.17} & \underline{90.80} & \underline{90.43} & 87.54 & \underline{90.05} & 82.54 & 87.97 & \textbf{88.81} & 77.89 & \underline{79.81} & 85.71 & \underline{87.63} \\
\hline
\multirow{10}{*}{TC} & MMANet~\citep{wei2023mmanet}\textsuperscript{a} & 86.63 & 86.39 & 85.68 & 74.48 & 85.87 & 85.21 & 72.65 & 71.88 & 72.73 & 80.99 & 85.24 & 65.79 & 57.47 & 74.05 & 70.21 & 76.99 \\
 & PASSION~\citep{shi2024passion}\textsuperscript{a} & 79.93 & 79.60 & 80.54 & 68.62 & 79.51 & 78.21 & 65.98 & 67.06 & 66.17 & 76.33 & 78.74 & 55.47 & 59.50 & 73.11 & 61.76 & 71.37 \\
 & DMAF-Net~\citep{lan2025dmaf}\textsuperscript{a} & 80.47 & 80.30 & 81.06 & 68.84 & 79.44 & 78.63 & 66.42 & 67.34 & 66.31 & 76.85 & 79.98 & 56.02 & 60.55 & 74.31 & 63.18 & 71.98 \\
 & IM-Fuse~\citep{pipoli2025imfuse}\textsuperscript{a} & 85.51 & 85.74 & 85.21 & 75.50 & 85.50 & 85.15 & 73.81 & 73.88 & 73.90 & 84.64 & 85.46 & 70.18 & 66.46 & 83.09 & 71.22 & 79.01 \\
 & IM-Fuse + MGML~\citep{zou2026mgml}\textsuperscript{a} & 85.46 & 85.52 & 84.34 & 75.82 & 85.41 & 85.17 & 74.66 & 73.90 & 74.70 & 84.84 & 85.78 & \underline{71.20} & 67.29 & 83.44 & 71.89 & 79.36 \\
 & LS$^3$M~\citep{zhang2025ls3m}\textsuperscript{b} & 85.86 & 85.83 & 85.84 & \underline{76.87} & 86.05 & 86.32 & \underline{76.87} & \underline{74.32} & \underline{75.25} & 84.09 & 85.18 & \textbf{71.98} & 66.67 & 83.61 & \underline{72.30} & \underline{79.76} \\
\cline{2-18}
 & RobustSeg~\citep{chen2019robustseg} & 83.04 & 83.02 & 82.05 & 70.60 & 82.93 & 82.01 & 68.34 & 66.59 & 68.32 & 80.44 & 82.60 & 60.84 & 54.41 & 76.83 & 62.00 & 73.60 \\
 & RFNet~\citep{ding2021rfnet} & 86.13 & 85.88 & 86.15 & 74.80 & 85.06 & 85.22 & 72.85 & 72.80 & 72.66 & 83.57 & 85.82 & 66.97 & 63.95 & 82.85 & 67.57 & 78.15 \\
 & mmFormer~\citep{zhang2022mmformer} & \underline{86.96} & \underline{86.93} & \underline{86.91} & 74.42 & \underline{86.71} & \underline{86.55} & 71.54 & 73.17 & 72.92 & \underline{86.60} & \underline{86.69} & 68.69 & \textbf{67.94} & \underline{86.16} & 67.61 & 79.32 \\
 & \textbf{Ours} & \textbf{88.61} & \textbf{88.41} & \textbf{88.18} & \textbf{78.42} & \textbf{88.44} & \textbf{87.61} & \textbf{76.91} & \textbf{76.68} & \textbf{75.41} & \textbf{87.44} & \textbf{87.90} & 70.95 & \underline{67.92} & \textbf{86.33} & \textbf{73.68} & \textbf{81.53} \\
\hline
\multirow{10}{*}{ET} & MMANet~\citep{wei2023mmanet}\textsuperscript{a} & 72.43 & 71.81 & 71.98 & 44.60 & 71.94 & 70.43 & 44.23 & 43.58 & 41.50 & 68.87 & 71.86 & 36.40 & 28.22 & 66.07 & 41.91 & 56.39 \\
 & PASSION~\citep{shi2024passion}\textsuperscript{a} & 69.77 & 68.97 & 69.90 & 42.94 & 69.27 & 69.58 & 40.66 & 40.85 & 38.09 & 68.27 & 67.86 & 28.47 & 31.25 & 66.71 & 37.12 & 53.92 \\
 & DMAF-Net~\citep{lan2025dmaf}\textsuperscript{a} & 69.45 & 69.06 & 70.11 & 43.35 & 69.50 & 70.37 & 41.66 & 40.33 & 38.20 & 68.62 & 68.02 & 30.37 & 31.64 & 66.97 & 36.86 & 54.30 \\
 & IM-Fuse~\citep{pipoli2025imfuse}\textsuperscript{a} & 76.63 & 77.00 & 76.29 & 50.03 & 76.68 & 74.81 & 49.09 & \underline{49.24} & 45.41 & 76.48 & 76.06 & 41.21 & 36.25 & 73.11 & 47.37 & 61.70 \\
 & IM-Fuse + MGML~\citep{zou2026mgml}\textsuperscript{a} & \underline{77.72} & \underline{79.48} & \underline{77.33} & \underline{51.68} & 75.56 & 76.11 & 49.17 & 49.01 & \underline{46.76} & \underline{77.84} & 77.25 & \underline{41.36} & \underline{37.77} & 73.90 & \textbf{49.23} & 62.67 \\
 & LS$^3$M~\citep{zhang2025ls3m}\textsuperscript{b} & \textbf{80.94} & \textbf{81.53} & \textbf{80.51} & 51.53 & \textbf{80.75} & \textbf{79.17} & \textbf{51.53} & 47.04 & 45.07 & \textbf{79.49} & \textbf{78.69} & 40.34 & 36.57 & \textbf{77.68} & 45.30 & \textbf{63.60} \\
\cline{2-18}
 & RobustSeg~\citep{chen2019robustseg} & 71.66 & 72.01 & 71.31 & 43.99 & 71.91 & 70.92 & 42.27 & 40.00 & 39.75 & 70.25 & 71.56 & 34.75 & 29.05 & 68.05 & 36.53 & 55.60 \\
 & RFNet~\citep{ding2021rfnet} & 74.19 & 74.04 & 74.01 & 44.43 & 73.89 & 73.82 & 42.99 & 41.35 & 40.33 & 72.42 & 74.05 & 34.32 & 32.07 & 71.33 & 38.29 & 57.44 \\
 & mmFormer~\citep{zhang2022mmformer} & 74.58 & 75.40 & 74.62 & 44.28 & 74.68 & 74.55 & 43.30 & 41.79 & 39.42 & 74.98 & 75.26 & 34.43 & 32.73 & 75.10 & 38.90 & 58.27 \\
 & \textbf{Ours} & 77.30 & 77.54 & 77.30 & \textbf{52.22} & \underline{77.16} & \underline{77.22} & \underline{50.60} & \textbf{50.50} & \textbf{48.60} & 76.86 & \underline{77.67} & \textbf{43.51} & \textbf{40.38} & \underline{76.76} & \underline{48.50} & \underline{63.47} \\
\hline
\multirow{10}{*}{Avg.} & MMANet~\citep{wei2023mmanet}\textsuperscript{a} & 82.77 & 81.53 & 82.34 & 69.02 & 81.91 & 80.79 & 68.12 & 66.93 & 67.29 & 73.61 & 80.91 & 61.80 & 50.01 & 68.19 & 65.00 & 72.01 \\
 & PASSION~\citep{shi2024passion}\textsuperscript{a} & 79.81 & 79.31 & 79.48 & 66.79 & 79.14 & 74.70 & 64.09 & 65.25 & 63.99 & 77.17 & 77.62 & 51.48 & 58.32 & 70.17 & 60.82 & 69.86 \\
 & DMAF-Net~\citep{lan2025dmaf}\textsuperscript{a} & 79.84 & 79.76 & 79.81 & 67.24 & 79.08 & 75.52 & 64.65 & 65.31 & 64.04 & 77.46 & 78.22 & 53.10 & 59.04 & 70.64 & 61.23 & 70.33 \\
 & IM-Fuse~\citep{pipoli2025imfuse}\textsuperscript{a} & 84.39 & 83.84 & 83.98 & 72.09 & 84.21 & 83.18 & 70.93 & \underline{70.38} & 69.71 & 80.80 & 83.18 & 66.31 & 60.08 & 77.70 & 68.29 & 75.93 \\
 & IM-Fuse + MGML~\citep{zou2026mgml}\textsuperscript{a} & 84.80 & 84.71 & 84.09 & 72.83 & 83.90 & 83.74 & 71.31 & 70.35 & \underline{70.46} & 81.65 & 83.83 & 66.93 & \underline{61.27} & 78.47 & \textbf{69.31} & 76.53 \\
 & LS$^3$M~\citep{zhang2025ls3m}\textsuperscript{b} & \textbf{86.32} & \textbf{85.70} & \textbf{86.07} & \underline{73.28} & \textbf{86.11} & \textbf{85.50} & \textbf{73.28} & 70.08 & 70.30 & \underline{82.05} & \underline{84.35} & \underline{66.99} & 60.85 & 79.84 & 68.30 & \underline{77.19} \\
\cline{2-18}
 & RobustSeg~\citep{chen2019robustseg} & 81.45 & 80.61 & 80.74 & 68.00 & 81.30 & 80.14 & 66.29 & 64.10 & 65.45 & 75.89 & 80.10 & 59.54 & 51.67 & 72.14 & 60.30 & 71.18 \\
 & RFNet~\citep{ding2021rfnet} & 83.71 & 82.62 & 83.62 & 69.82 & 83.08 & 82.98 & 68.55 & 67.10 & 67.34 & 79.07 & 82.44 & 62.75 & 57.85 & 77.67 & 63.34 & 74.13 \\
 & mmFormer~\citep{zhang2022mmformer} & 84.23 & 83.61 & 84.18 & 69.74 & 84.07 & 83.89 & 68.30 & 67.46 & 67.36 & 81.25 & 83.32 & 63.84 & 60.16 & \underline{80.42} & 64.12 & 75.06 \\
 & \textbf{Ours} & \underline{85.74} & \underline{84.79} & \underline{85.55} & \textbf{73.81} & \underline{85.59} & \underline{85.21} & \underline{72.65} & \textbf{71.57} & \textbf{71.35} & \textbf{82.28} & \textbf{84.51} & \textbf{67.76} & \textbf{62.07} & \textbf{80.96} & \underline{69.30} & \textbf{77.54} \\
\hline
\end{tabular}%
}}
\par\vspace{1pt}
\begin{minipage}{\linewidth}
\fontsize{8}{9}\selectfont
\ModPresent\ = available modality; \ModAbsent\ = missing modality. M denotes the tumor-region metric. Unmarked rows: this study. \textsuperscript{a}Results from Zou et~al.~\citep{zou2026mgml}. \textsuperscript{b}Results from Zhang et~al.~\citep{zhang2025ls3m}. References beside method names identify the original methods. Bold marks the highest score and underlining the second-highest distinct score among all methods shown. Ties receive the same marking. Avg. is the arithmetic mean of WT, TC, and ET for each subset. Published WT, TC, and ET means are reproduced from the cited sources.
\end{minipage}
\end{table}
\end{landscape}
\restoregeometry
\FloatBarrier

\subsubsection{BraTS2023 GLI}

On BraTS2023 GLI, PMTM obtains an average Dice of 80.47\%,
2.78 percentage points higher than mmFormer.
WT Dice is higher by 0.29 points, TC Dice by 3.30 points,
and ET Dice by 4.75 points in our experiments.

Table~\ref{tab:brats2023_published_comparison} also includes
the BraTS2023 results reported in the experimental study of
Zou et~al.~\citep{zou2026mgml}.
The published rows report IM-Fuse and IM-Fuse + MGML
as evaluated in that study.
PMTM has an average Dice of 80.47\%, compared with 80.28\%
for IM-Fuse and 81.26\% for IM-Fuse + MGML in that study.
PMTM has higher ET Dice than these two reported results,
but IM-Fuse + MGML has higher WT and TC Dice.

\begin{table}[!htbp]
\centering
\small
\setlength{\tabcolsep}{4pt}
\caption{BraTS2023 GLI Dice (\%) averaged over all 15 non-empty modality subsets. The Source column identifies experiments conducted in this study and published results. Bold marks the highest score and underlining the second-highest distinct score among all methods shown. Ties receive the same marking.}
\label{tab:brats2023_published_comparison}
\begin{tabular}{llrrrr}
\toprule
Method & Source & WT $\uparrow$ & TC $\uparrow$ & ET $\uparrow$ & Avg. $\uparrow$ \\
\midrule
IM-Fuse \citep{pipoli2025imfuse} & \citep{zou2026mgml} & \underline{89.69} & 81.74 & 69.41 & 80.28 \\
IM-Fuse + MGML \citep{zou2026mgml} & \citep{zou2026mgml} & \textbf{90.47} & \textbf{83.01} & \underline{70.30} & \textbf{81.26} \\
\midrule
RobustSeg \citep{chen2019robustseg} & This study & 84.69 & 73.10 & 60.69 & 72.82 \\
RFNet \citep{ding2021rfnet} & This study & 85.68 & 75.66 & 61.71 & 74.35 \\
mmFormer \citep{zhang2022mmformer} & This study & 87.45 & 79.42 & 66.21 & 77.69 \\
\textbf{Ours} & This study & 87.74 & \underline{82.72} & \textbf{70.96} & \underline{80.47} \\
\bottomrule
\end{tabular}
\end{table}

Table~\ref{tab:brats2023_rfnet_style_missing_methods}
reports the local and published results for individual modality
subsets, grouped by WT, TC, ET, and three-region average.
Among the four methods evaluated in this study, PMTM has
the highest three-region average (Avg.) in all 15 subsets.
It also has the highest TC and ET Dice in all 15 subsets
within this group.
Its WT Dice is the highest in 12 subsets.
mmFormer has the highest WT Dice for FLAIR+T1, FLAIR-only,
and T1ce-only input.

\newgeometry{left=24pt,right=24pt,top=50pt,bottom=50pt}
\begin{landscape}
\begin{table}[p]
\centering
\captionsetup{font=small}
\caption{BraTS2023 GLI Dice (\%) for all 15 non-empty modality subsets.}
\label{tab:brats2023_rfnet_style_missing_methods}
\setlength{\arrayrulewidth}{0.25pt}
\setlength{\tabcolsep}{0.9pt}
\renewcommand{\arraystretch}{1.0}
\fontsize{8}{8.4}\selectfont
\makebox[\linewidth][c]{\resizebox{0.96\linewidth}{!}{%
\begin{tabular}{l|l|r|r|r|r|r|r|r|r|r|r|r|r|r|r|r|r}
\hline
\multirow{4}{*}{M} & \multicolumn{1}{c|}{FLAIR} & \multicolumn{1}{c|}{\ModPresent} & \multicolumn{1}{c|}{\ModAbsent} & \multicolumn{1}{c|}{\ModPresent} & \multicolumn{1}{c|}{\ModPresent} & \multicolumn{1}{c|}{\ModPresent} & \multicolumn{1}{c|}{\ModPresent} & \multicolumn{1}{c|}{\ModPresent} & \multicolumn{1}{c|}{\ModAbsent} & \multicolumn{1}{c|}{\ModPresent} & \multicolumn{1}{c|}{\ModAbsent} & \multicolumn{1}{c|}{\ModAbsent} & \multicolumn{1}{c|}{\ModPresent} & \multicolumn{1}{c|}{\ModAbsent} & \multicolumn{1}{c|}{\ModAbsent} & \multicolumn{1}{c|}{\ModAbsent} & \multirow{4}{*}{Mean} \\
 & \multicolumn{1}{c|}{T1ce} & \multicolumn{1}{c|}{\ModPresent} & \multicolumn{1}{c|}{\ModPresent} & \multicolumn{1}{c|}{\ModPresent} & \multicolumn{1}{c|}{\ModAbsent} & \multicolumn{1}{c|}{\ModPresent} & \multicolumn{1}{c|}{\ModPresent} & \multicolumn{1}{c|}{\ModAbsent} & \multicolumn{1}{c|}{\ModAbsent} & \multicolumn{1}{c|}{\ModAbsent} & \multicolumn{1}{c|}{\ModPresent} & \multicolumn{1}{c|}{\ModPresent} & \multicolumn{1}{c|}{\ModAbsent} & \multicolumn{1}{c|}{\ModAbsent} & \multicolumn{1}{c|}{\ModPresent} & \multicolumn{1}{c|}{\ModAbsent} &  \\
 & \multicolumn{1}{c|}{T1} & \multicolumn{1}{c|}{\ModPresent} & \multicolumn{1}{c|}{\ModPresent} & \multicolumn{1}{c|}{\ModAbsent} & \multicolumn{1}{c|}{\ModPresent} & \multicolumn{1}{c|}{\ModPresent} & \multicolumn{1}{c|}{\ModAbsent} & \multicolumn{1}{c|}{\ModAbsent} & \multicolumn{1}{c|}{\ModPresent} & \multicolumn{1}{c|}{\ModPresent} & \multicolumn{1}{c|}{\ModPresent} & \multicolumn{1}{c|}{\ModAbsent} & \multicolumn{1}{c|}{\ModAbsent} & \multicolumn{1}{c|}{\ModPresent} & \multicolumn{1}{c|}{\ModAbsent} & \multicolumn{1}{c|}{\ModAbsent} &  \\
 & \multicolumn{1}{c|}{T2} & \multicolumn{1}{c|}{\ModPresent} & \multicolumn{1}{c|}{\ModPresent} & \multicolumn{1}{c|}{\ModPresent} & \multicolumn{1}{c|}{\ModPresent} & \multicolumn{1}{c|}{\ModAbsent} & \multicolumn{1}{c|}{\ModAbsent} & \multicolumn{1}{c|}{\ModPresent} & \multicolumn{1}{c|}{\ModPresent} & \multicolumn{1}{c|}{\ModAbsent} & \multicolumn{1}{c|}{\ModAbsent} & \multicolumn{1}{c|}{\ModPresent} & \multicolumn{1}{c|}{\ModAbsent} & \multicolumn{1}{c|}{\ModAbsent} & \multicolumn{1}{c|}{\ModAbsent} & \multicolumn{1}{c|}{\ModPresent} &  \\
\hline
\multirow{6}{*}{WT} & IM-Fuse~\citep{pipoli2025imfuse}\textsuperscript{a} & \underline{93.62} & \underline{91.46} & \underline{93.11} & \underline{92.55} & \underline{92.73} & \underline{91.33} & \underline{92.15} & \underline{89.47} & \underline{91.74} & \underline{84.76} & \underline{89.89} & 89.48 & \underline{82.01} & \underline{82.22} & \underline{88.94} & \underline{89.69} \\
 & IM-Fuse + MGML~\citep{zou2026mgml}\textsuperscript{a} & \textbf{94.45} & \textbf{92.12} & \textbf{93.81} & \textbf{93.34} & \textbf{93.43} & \textbf{91.92} & \textbf{92.71} & \textbf{90.49} & \textbf{92.48} & \textbf{85.73} & \textbf{90.81} & \textbf{90.95} & \textbf{82.60} & \textbf{82.83} & \textbf{89.42} & \textbf{90.47} \\
\cline{2-18}
 & RobustSeg~\citep{chen2019robustseg} & 89.40 & 86.00 & 89.20 & 88.70 & 88.90 & 88.50 & 88.10 & 84.70 & 87.90 & 77.30 & 86.60 & 86.00 & 72.50 & 73.80 & 82.70 & 84.69 \\
 & RFNet~\citep{ding2021rfnet} & 89.77 & 86.89 & 89.78 & 89.20 & 88.65 & 88.33 & 89.00 & 86.39 & 87.65 & 79.69 & 86.58 & 86.01 & 76.21 & 76.35 & 84.70 & 85.68 \\
 & mmFormer~\citep{zhang2022mmformer} & 91.21 & 87.62 & 91.12 & 90.89 & 90.93 & 90.75 & 90.60 & 87.16 & 90.65 & 81.33 & 87.35 & \underline{89.57} & 77.84 & 79.04 & 85.68 & 87.45 \\
 & \textbf{Ours} & 91.74 & 88.18 & 91.67 & 91.24 & 91.21 & 90.98 & 91.06 & 87.36 & 90.47 & 81.54 & 87.90 & 89.24 & 78.21 & 78.93 & 86.31 & 87.74 \\
\hline
\multirow{6}{*}{TC} & IM-Fuse~\citep{pipoli2025imfuse}\textsuperscript{a} & 89.54 & 88.22 & 87.31 & 78.64 & 87.77 & 87.22 & 76.65 & 76.02 & 77.03 & 86.64 & 87.93 & \underline{74.91} & 69.31 & 84.88 & \underline{74.01} & 81.74 \\
 & IM-Fuse + MGML~\citep{zou2026mgml}\textsuperscript{a} & \textbf{90.02} & \textbf{89.73} & \underline{88.95} & \textbf{79.33} & \underline{89.14} & \underline{88.43} & \textbf{78.25} & \textbf{77.65} & \textbf{78.61} & \underline{88.09} & \underline{88.63} & \textbf{75.83} & \underline{70.22} & \underline{86.73} & \textbf{75.54} & \textbf{83.01} \\
\cline{2-18}
 & RobustSeg~\citep{chen2019robustseg} & 81.80 & 80.90 & 81.30 & 69.80 & 81.60 & 80.90 & 67.70 & 65.70 & 68.00 & 79.10 & 79.70 & 62.10 & 58.90 & 77.30 & 61.70 & 73.10 \\
 & RFNet~\citep{ding2021rfnet} & 83.93 & 84.27 & 83.67 & 72.08 & 82.96 & 82.18 & 70.48 & 70.21 & 69.53 & 81.38 & 83.76 & 65.21 & 61.38 & 78.80 & 65.05 & 75.66 \\
 & mmFormer~\citep{zhang2022mmformer} & 87.82 & 87.44 & 87.53 & 74.66 & 87.44 & 86.87 & 73.92 & 71.05 & 72.82 & 86.69 & 86.77 & 69.78 & 65.49 & 84.96 & 67.99 & 79.42 \\
 & \textbf{Ours} & \underline{89.65} & \underline{89.23} & \textbf{89.39} & \underline{78.99} & \textbf{89.34} & \textbf{88.77} & \underline{77.62} & \underline{76.36} & \underline{77.69} & \textbf{88.28} & \textbf{88.74} & 73.88 & \textbf{72.21} & \textbf{87.26} & 73.45 & \underline{82.72} \\
\hline
\multirow{6}{*}{ET} & IM-Fuse~\citep{pipoli2025imfuse}\textsuperscript{a} & \underline{85.41} & \underline{84.64} & \underline{83.81} & 58.83 & 82.66 & 81.45 & 55.88 & 54.12 & 52.39 & \underline{84.57} & \underline{84.91} & 49.67 & 44.26 & 81.65 & \underline{56.93} & 69.41 \\
 & IM-Fuse + MGML~\citep{zou2026mgml}\textsuperscript{a} & \textbf{86.21} & \textbf{85.46} & \textbf{84.39} & \underline{59.53} & \underline{83.18} & \underline{82.41} & \underline{56.83} & \underline{55.27} & \underline{53.36} & \textbf{85.54} & \textbf{85.73} & \underline{51.46} & \underline{45.02} & \textbf{82.59} & \textbf{57.53} & \underline{70.30} \\
\cline{2-18}
 & RobustSeg~\citep{chen2019robustseg} & 75.10 & 74.40 & 74.30 & 51.20 & 75.20 & 74.20 & 49.70 & 46.30 & 47.70 & 73.20 & 73.90 & 42.80 & 37.50 & 71.80 & 43.00 & 60.69 \\
 & RFNet~\citep{ding2021rfnet} & 75.95 & 76.30 & 75.45 & 53.20 & 75.06 & 74.02 & 51.43 & 50.41 & 47.87 & 73.25 & 75.48 & 42.70 & 37.77 & 70.64 & 46.06 & 61.71 \\
 & mmFormer~\citep{zhang2022mmformer} & 80.72 & 80.51 & 80.37 & 56.12 & 80.64 & 79.84 & 54.59 & 51.41 & 52.36 & 79.75 & 79.92 & 47.48 & 43.07 & 78.34 & 48.07 & 66.21 \\
 & \textbf{Ours} & 83.08 & 83.11 & 82.86 & \textbf{61.67} & \textbf{83.19} & \textbf{82.76} & \textbf{60.20} & \textbf{58.98} & \textbf{59.13} & 82.32 & 82.80 & \textbf{54.16} & \textbf{52.53} & \underline{81.95} & 55.64 & \textbf{70.96} \\
\hline
\multirow{6}{*}{Avg.} & IM-Fuse~\citep{pipoli2025imfuse}\textsuperscript{a} & \underline{89.52} & \underline{88.11} & \underline{88.08} & 76.67 & 87.72 & 86.67 & 74.89 & 73.20 & 73.72 & \underline{85.32} & \underline{87.58} & 71.35 & 65.19 & \underline{82.92} & \underline{73.29} & 80.28 \\
 & IM-Fuse + MGML~\citep{zou2026mgml}\textsuperscript{a} & \textbf{90.23} & \textbf{89.10} & \textbf{89.05} & \textbf{77.40} & \textbf{88.58} & \textbf{87.59} & \underline{75.93} & \textbf{74.47} & \underline{74.82} & \textbf{86.45} & \textbf{88.39} & \textbf{72.75} & \underline{65.95} & \textbf{84.05} & \textbf{74.16} & \textbf{81.26} \\
\cline{2-18}
 & RobustSeg~\citep{chen2019robustseg} & 82.10 & 80.43 & 81.60 & 69.90 & 81.90 & 81.20 & 68.50 & 65.57 & 67.87 & 76.53 & 80.07 & 63.63 & 56.30 & 74.30 & 62.47 & 72.82 \\
 & RFNet~\citep{ding2021rfnet} & 83.22 & 82.49 & 82.97 & 71.49 & 82.22 & 81.51 & 70.30 & 69.00 & 68.35 & 78.11 & 81.94 & 64.64 & 58.45 & 75.26 & 65.27 & 74.35 \\
 & mmFormer~\citep{zhang2022mmformer} & 86.58 & 85.19 & 86.34 & 73.89 & 86.34 & 85.82 & 73.04 & 69.87 & 71.94 & 82.59 & 84.68 & 68.94 & 62.13 & 80.78 & 67.25 & 77.69 \\
 & \textbf{Ours} & 88.16 & 86.84 & 87.97 & \underline{77.30} & \underline{87.91} & \underline{87.50} & \textbf{76.29} & \underline{74.23} & \textbf{75.76} & 84.05 & 86.48 & \underline{72.43} & \textbf{67.65} & 82.72 & 71.80 & \underline{80.47} \\
\hline
\end{tabular}%
}}
\par\vspace{1pt}
\begin{minipage}{\linewidth}
\fontsize{8}{9}\selectfont
\ModPresent\ = available modality; \ModAbsent\ = missing modality. M denotes the tumor-region metric. Unmarked rows: this study. \textsuperscript{a}Results from Zou et~al.~\citep{zou2026mgml}. References beside method names identify the original methods. Bold marks the highest score and underlining the second-highest distinct score among all methods shown. Ties receive the same marking. Avg. is the arithmetic mean of WT, TC, and ET for each subset. Mean retains the source-reported regional averages and their three-region mean.
\end{minipage}
\end{table}
\end{landscape}
\restoregeometry
\FloatBarrier

\subsubsection{Effect of modality availability on BraTS2020}

We next examine how modality availability affects PMTM
on BraTS2020.
We group the 15 subsets in
Table~\ref{tab:rfnet_style_missing_methods}
by the number of available modalities.
Average Dice increases from 70.02\% with one modality
to 77.93\% with two, 82.43\% with three, and 85.74\%
with all four.
These averages cover different modality combinations.
They do not imply that adding any sequence improves
every regional score.

To examine the absence of a specific sequence, we compare
complete input with input containing the other three sequences.
Without T1ce, the available modalities are T1, FLAIR, and T2.
TC Dice decreases from 88.61\% to 78.42\%,
a drop of 10.19 percentage points.
ET Dice decreases from 77.30\% to 52.22\%,
a drop of 25.08 percentage points.
Average Dice decreases from 85.74\% to 73.81\%.

Without FLAIR, the available modalities are T1ce, T1, and T2.
WT Dice decreases from 91.33\% to 88.41\%,
a drop of 2.92 percentage points.
Average Dice decreases from 85.74\% to 84.79\%.
ET Dice increases slightly from 77.30\% to 77.54\%.

We also group the incomplete subsets by whether T1ce is available.
Average Dice is 84.13\% across the seven subsets containing
T1ce and 69.79\% across the seven subsets without T1ce.
The corresponding ET Dice values are 77.22\% and 47.76\%.
These groups also differ in the number and identity of
the other available sequences.
Their difference therefore does not isolate the effect
of removing T1ce.

\subsection{Component ablation on BraTS2020}

We evaluate four variants to examine the tumor-state path,
cross-stage state propagation, semantic assignment supervision,
and modality-token interaction.
All variants use the same BraTS2020 subject-level partitions
and evaluation protocol as the main comparison.

The variant without the tumor-state path removes it from all four stages.
It combines the available modality features by averaging and feature projection.
The segmentation loss and auxiliary supervision remain unchanged.
We use the same training settings and budget as for the full PMTM model.

The local state memory variant retains state construction and
state-guided fusion. It reinitializes the state at every stage,
removing propagation between stages.

The remaining two variants remove semantic assignment supervision
at all four stages or remove modality-token interaction from all
fusion stages. Removing semantic supervision changes only the
training objective, not the network architecture.

\begin{table*}[t]
\centering
\caption{Ablation results on BraTS2020, averaged over 100 subjects
and all 15 modality subsets. Local state memory removes cross-stage
state propagation. Dice values are percentages. Bold marks the highest score
and underlining the second-highest distinct score among the variants.}
\label{tab:ablation_ours}
\resizebox{\textwidth}{!}{%
\begin{tabular}{lcccc}
\toprule
Variant & WT & TC & ET & Avg. \\
\midrule

w/o tumor-state path
 & 86.74 & 78.44 & 59.02 & 74.73 \\

Local state memory
 & 87.31 & 80.49 & 62.08 & 76.63 \\

w/o semantic assignment supervision
 & 87.42 & 78.85 & 60.29 & 75.52 \\

w/o modality-token interaction
 & \underline{87.58} & \underline{81.25} & \underline{63.01} & \underline{77.28} \\

\midrule
\textbf{Ours}
 & \textbf{87.63} & \textbf{81.53} & \textbf{63.47}
 & \textbf{77.54} \\
\bottomrule
\end{tabular}%
}
\end{table*}

Table~\ref{tab:ablation_ours} reports an average Dice
of 77.54\% for PMTM with all components, matching
Table~\ref{tab:main_benchmark_all}.
Removing the tumor-state path reduces average Dice
to 74.73\%, a decrease of 2.81 percentage points.
WT Dice decreases by 0.89 points, TC by 3.09 points,
and ET by 4.45 points.

Reinitializing the state at each stage reduces average Dice
by 0.91 percentage points.
WT Dice decreases by 0.32 points, TC by 1.04 points,
and ET by 1.39 points.

Removing semantic assignment supervision reduces average Dice
by 2.02 percentage points.
WT Dice decreases by 0.21 points, TC by 2.68 points,
and ET by 3.18 points.

Removing modality-token interaction reduces average Dice
by 0.26 percentage points.
WT Dice decreases by 0.05 points, TC by 0.28 points,
and ET by 0.46 points.

\subsection{Semantic assignment and tumor-state dynamics}
\label{sec:state_analysis}

The ablations measure changes in segmentation performance
when components are removed. We next examine the assignment
maps and state slots to describe the representations learned
by PMTM across encoder stages.

\subsubsection{Analysis protocol}
\label{sec:state_protocol}

We use the selected checkpoint to perform the internal state
analysis on all 50 subjects in the BraTS2020 validation partition.
The analysis includes complete input, FLAIR-only input,
and T2-only input.
State statistics are computed only for sliding windows
containing at least one ground-truth tumor voxel.

At each of the four stages, we upsample the assignment maps
using trilinear interpolation.
We then average them over the overlapping windows used
for segmentation.
Applying argmax at each voxel gives hard assignment labels.
WT combines NCR/NET, ED, and ET.
TC combines NCR/NET and ET, while ET uses the enhancing class.
We compute Dice for these three regions.
We also measure the mean activation of each assignment channel
for voxels belonging to each ground-truth class.
Normalized entropy and mean maximum assignment probability
describe how concentrated the assignments are.

We measure mean cosine similarity between different state slots
within each stage.
To examine slot correspondence across stages, we use stage~1
as the reference for stages~2--4.
We report diagonal matching and a similarity margin.
This margin is the similarity to the corresponding reference
slot minus the largest similarity to a different reference slot.
We also measure the relative magnitude of state updates.
Finally, we compare corresponding slots obtained with complete
and incomplete input using cosine similarity.

\subsubsection{Assignment maps and state diagnostics}

\begin{figure}[!htbp]
\centering
\includegraphics[width=0.95\textwidth]{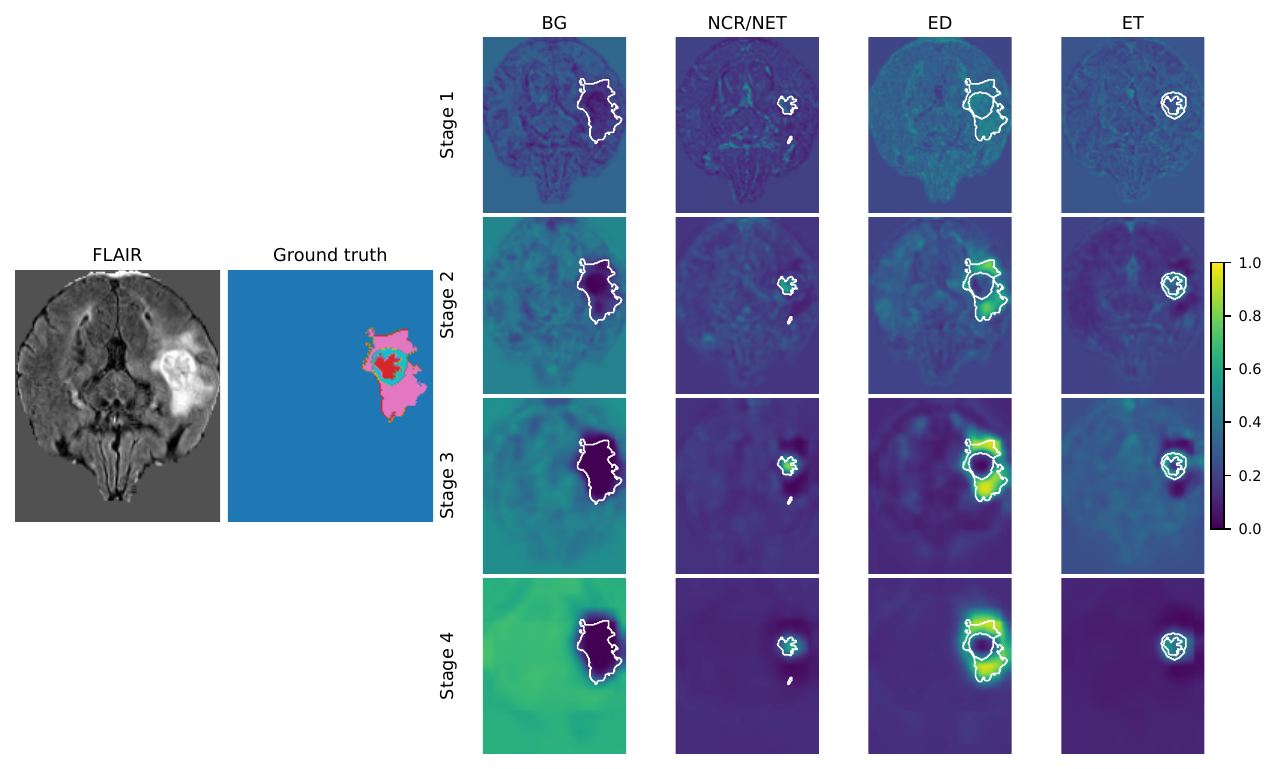}
\caption{Assignment probabilities at encoder stages~1--4
for a representative BraTS2020 subject under complete input
(axial slice $z=61$).
The FLAIR image and ground-truth segmentation are shown
for reference. White contours mark the boundary of the
corresponding ground-truth channel. The ordered assignment
channels follow BG, NCR/NET, ED, and ET.}
\label{fig:assignment_maps}
\end{figure}

Figure~\ref{fig:assignment_maps} shows assignment probabilities
for a representative BraTS2020 subject under complete input.
The stage~1 maps retain anatomical texture and are more diffuse
than the later maps.
The stage~4 responses are concentrated around the tumor.
At the deeper stages, the BG channel is suppressed within
the lesion, while the ED channel has a broader response.
The NCR/NET and ET channels overlap within the tumor core.
These maps show the internal assignments, not the final
decoder segmentation.

Table~\ref{tab:state_diagnostics} reports the quantitative
diagnostics under complete input.
Average assignment Dice increases from \StateOneDice\%
at stage~1 to \StateFourDice\% at stage~4.
The averages at stages~2 and~3 are similar.
ET Dice decreases from 37.12\% at stage~2 to 25.96\%
at stage~3, then reaches 41.50\% at stage~4.

\begin{table*}[!htbp]
\centering
\caption{Internal state diagnostics on all 50 subjects in the
BraTS2020 validation partition under complete input.
Assignment Dice values are percentages. For these Dice columns, bold
and underlining mark the highest and second-highest distinct scores
across stages.
Identity matching uses the corresponding stage~1 slot
as the reference.}
\label{tab:state_diagnostics}
\resizebox{\textwidth}{!}{%
\begin{tabular}{ccccccccc}
\toprule
Stage & WT & TC & ET & Avg. & Norm. entropy & Mean max probability & \shortstack{Mean cosine similarity\\between different state slots} & Identity match \\
\midrule
1 & 15.22 & 8.18 & 7.36 & 10.25 & 0.974 & 0.334 & -0.286 & -- \\
2 & 48.72 & 32.93 & \underline{37.12} & 39.59 & 0.939 & 0.416 & -0.215 & 1.000 \\
3 & \underline{55.85} & \underline{37.79} & 25.96 & \underline{39.87} & 0.883 & 0.461 & -0.049 & 1.000 \\
4 & \textbf{70.34} & \textbf{64.40} & \textbf{41.50} & \textbf{58.75} & 0.729 & 0.648 & 0.189 & 1.000 \\
\bottomrule
\end{tabular}%
}
\end{table*}

At stage~4, each ground-truth class has its highest conditional
mean activation in the intended assignment channel.
The margin over the next-highest channel is 0.47 for BG,
0.09 for NCR/NET, 0.52 for ED, and 0.04 for ET.
The channels therefore show a preference for their intended
classes, but their responses are not exclusive to those classes.
The separation is smaller for NCR/NET and ET.

The assignments become more concentrated across stages.
Normalized entropy decreases from \StateOneEntropy\
at stage~1 to \StateFourEntropy\ at stage~4.
Mean maximum assignment probability increases from
\StateOneMaxP\ to \StateFourMaxP\ over the same stages.

Mean cosine similarity between different state slots increases
from \StateOneTokenCos\ at stage~1 to \StateFourTokenCos\
at stage~4.
Identity matching relative to stage~1 is reported as 1.000
at stages~2--4.
The corresponding similarity margins are
\StateTwoMargin, \StateThreeMargin, and \StateFourMargin.
The slots become more similar with depth, but they retain
their correspondence to the stage~1 slots.

Under complete input, relative state-update magnitudes
range from 0.20 to 0.26 across slots and stages.
For FLAIR-only and T2-only input, mean cosine similarity
to the corresponding complete-input slots ranges from
approximately 0.9968 to 0.9999 across slots and stages.

\subsection{Qualitative segmentation comparison}

\begin{figure}[!htbp]
\centering
\includegraphics[width=\textwidth]{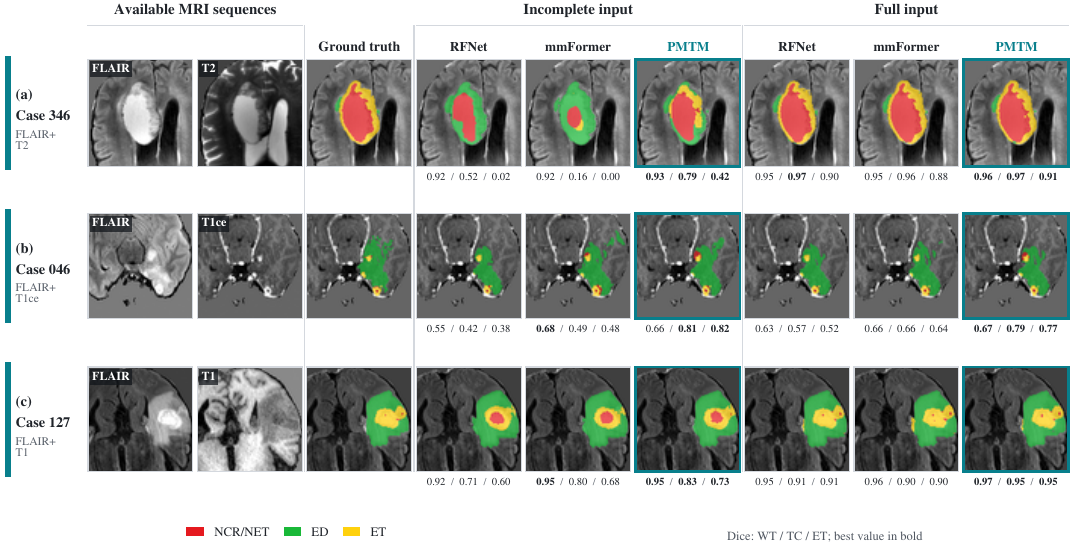}
\caption{Segmentation maps for three BraTS2020 test subjects
selected for illustration under incomplete and complete input.
Cases 346, 046, and 127 refer to BraTS20\_Training\_346,
BraTS20\_Training\_046, and BraTS20\_Training\_127.
The incomplete subsets are FLAIR+T2 for case 346,
T1ce+FLAIR for case 046, and T1+FLAIR for case 127.
Red, green, and yellow denote NCR/NET, ED, and ET.
Dice values below each prediction are reported on a
zero-to-one scale in WT/TC/ET order.
The highest displayed values within each input setting
are shown in bold, including ties.}
\label{fig:qualitative}
\end{figure}

Figure~\ref{fig:qualitative} compares PMTM with RFNet and
mmFormer for three BraTS2020 test subjects under complete
and incomplete input.
These cases are selected for illustration.

For case 346 with FLAIR+T2, RFNet and mmFormer predict
little of the enhancing region.
The PMTM prediction more closely matches the ground truth
in TC and ET.

For case 046 with T1ce+FLAIR, PMTM obtains higher TC and
ET Dice than mmFormer.
Its WT Dice is slightly lower than that of mmFormer.

For case 127 with T1+FLAIR, PMTM obtains the highest
TC and ET Dice among the three methods.
PMTM and mmFormer both have a displayed WT Dice of 0.95.

With complete input, the predictions are generally
more similar in these examples.
Visible differences in TC and ET remain for case 046.

\section{Discussion}
\label{sec:discussion}

In our experiments, the gains over mmFormer are larger
for TC and ET than for WT.
However, segmentation performance still depends on the
available MRI sequences.
In BraTS2020, removing T1ce from the complete input mainly
reduces TC and ET Dice.
Removing FLAIR mainly reduces WT Dice.
These observations are consistent with the different
information supplied by the sequences
\citep{ding2021rfnet,xiao2025mpdfunet}.
The memory can combine evidence from the remaining inputs,
but the results do not establish that it recovers all
information from a missing sequence.

The local state memory ablation retains state construction and
state-guided fusion but removes propagation between stages.
Its lower Dice suggests that evidence accumulated at earlier
stages contributes to segmentation within this architecture.
Removing semantic assignment supervision produces a larger
decrease than using local state memory.
Together, these comparisons support progressive state updates
and supervised assignments in the evaluated BraTS2020 configuration.
They do not establish how these components would interact
in other architectures or under different training settings.

The assignment channels show preferences for their intended
classes, but the NCR/NET and ET responses overlap.
This overlap limits their interpretation as exclusive
representations of individual tissue classes.
Preserved slot correspondence across stages is consistent
with an ordered state representation.
The high cosine similarity between states obtained with
complete input and with FLAIR-only or T2-only input indicates
similar vector directions.
It does not establish equivalent information content
or segmentation performance.

\subsection{Limitations}

We evaluate PMTM using fixed subject-level splits within
three BraTS cohorts.
The experiments do not measure sensitivity to alternative
splits or establish performance on external imaging protocols.
We do not report variability across training seeds.
Our experiments compare PMTM with RobustSeg, RFNet, and mmFormer.
Additional methods are compared using published scores;
these methods were not retrained in our experimental setup.

RobustSeg and RFNet use $80^3$ crops, while mmFormer and PMTM
use $128^3$ crops.
The comparisons with RobustSeg and RFNet therefore do not
separate differences in architecture from differences
in input crop size.
The comparison with mmFormer uses the same crop size.

Internal state analysis includes all 50 BraTS2020 validation
subjects under complete, FLAIR-only, and T2-only input.
The observed assignment preferences and slot correspondence
should be interpreted within these settings.
The experiments evaluate missing-modality subsets.
They do not establish performance under other acquisition
failures or on other tumor types.

\section{Conclusion}

We developed PMTM for brain tumor segmentation from
incomplete multimodal MRI.
The method accumulates tumor evidence across encoder stages
and uses the updated state to guide modality fusion.
We evaluated PMTM on BraTS2018, BraTS2020, and BraTS2023 GLI
across all 15 non-empty modality subsets.
Dice averaged over the three tumor regions and all 15 subsets
is 75.31\% on BraTS2018, 77.54\% on BraTS2020,
and 80.47\% on BraTS2023 GLI.
These averages are higher than those of RobustSeg, RFNet,
and mmFormer in our experiments on each dataset.

The BraTS2020 ablations support the contributions of the
tumor-state path, cross-stage propagation, and semantic
assignment supervision within the evaluated configuration.
Removing modality-token interaction gives the smallest
average Dice decrease among the tested variants,
at 0.26 percentage points.
Internal analysis shows that assignments become more
concentrated across stages and that state slots retain
their correspondence.
The NCR/NET and ET assignment responses still overlap.

\section*{Declaration of competing interest}
The authors declare that they have no known competing financial interests or personal relationships that could have appeared to influence the work reported in this paper.

\section*{Funding}
This research received no specific grant from funding agencies in the public, commercial, or not-for-profit sectors.

\section*{Ethics and data use}
This study used existing MRI data and segmentation annotations from the
BraTS collections cited in Section~\ref{sec:experiments}. No new participant
recruitment or image acquisition was performed for this study. Access and
reuse are governed by the terms of the respective data providers.

\section*{Data availability}
The BraTS datasets are available through the challenge and data-access
procedures of the collections cited in Section~\ref{sec:experiments}.
The PMTM source code and fixed subject-level split identifiers are available
at \url{https://github.com/milad67/PMTM}.
Access to the original MRI data remains subject to the terms of the
respective data providers.

\end{document}